\documentclass[%
reprint,
longbibliography,
amsmath,amssymb,
aps,
prfluids,
onecolumn,
notitlepage,
floatfix
]{revtex4-2}

\usepackage{graphicx}
\graphicspath{{./Figures/}}

\usepackage[labelformat=simple]{subcaption}
\usepackage{siunitx} 

\usepackage{empheq}
\usepackage{bm}      
\usepackage{mathrsfs}
\usepackage{cancel} 

\newcommand{\ri}{\mathrm{i}}
\newcommand{\re}{\mathrm{e}}
\DeclareMathOperator{\tr}{tr}
\DeclareMathOperator{\Real}{Re}
\DeclareMathOperator{\Imag}{Im}

\DeclareMathOperator{\symm}{symm}
\newcommand\Wor{{\text{Wo}}}  

\allowdisplaybreaks

\usepackage[colorlinks,citecolor=blue]{hyperref}

\begin{document}


\title{Theory and simulation of elastoinertial rectification of oscillatory flows in two-dimensional deformable rectangular channels}

\author{Uday M.\ Rade}
\author{Shrihari D.\ Pande}
\author{Ivan C.\ Christov}
\thanks{Corresponding author}
\email{christov@purdue.edu}\homepage{http://tmnt-lab.org}
\affiliation{School of Mechanical Engineering, Purdue University, West Lafayette, Indiana 47907, USA}

\date{\today}

\begin{abstract}
Oscillatory flows in compliant confinements underpin processes ranging from physiological transport in blood vessels and airways to flow control and pumping in soft microfluidic devices. To understand the fundamental physics behind such processes, we study how hydrodynamic forces induce deformation at the fluid--solid interface in a slender two-dimensional (2D) channel bounded below by a rigid bottom surface and above by a slender elastic layer. The nonlinear coupling between flow and deformation, along with the attendant geometric asymmetry caused by flow-induced deformation, produces a streaming effect (a nonzero cycle-average despite time-periodic forcing). Surprisingly, flow inertia provides another nonlinear coupling, tightly connected to deformation, that enhances streaming, termed ``elastoinertial rectification'' by Zhang and Rallabandi [J.\ Fluid Mech.\ \textbf{996}, A16 (2024)]. We adapt the latter theory of how two-way coupled fluid--structure interaction (FSI) produces streaming to a 2D rectangular configuration, specifically taking care to capture the deformations of the nearly incompressible slender elastic layer via the combined foundation model of Chandler and Vella [Proc.\ R.\ Soc.\ A \textbf{476}, 20200551 (2020)]. We put this elastoinertial rectification theory to a stringent test against direct numerical simulations performed using a stabilized, conforming arbitrary Lagrangian--Eulerian (ALE) FSI formulation, implemented via the open-source computing platform FEniCS. We examine the axial variation of the cycle-averaged pressure as a function of key dimensionless groups of the problem: the Womersley number and the elastoviscous number. Assuming a small compliance number, we find excellent agreement between a perturbative calculation based on elastoinertial rectification theory and the simulations for both the leading-order and cycle-averaged pressure and deformation across a range of conditions. Finally, unlike previous literature, the combined foundation model also predicts nontrivial horizontal displacements in confined 2D layers, consistent with the simulations.
\end{abstract}

\maketitle


\section{Introduction}
\label{sec:intro}

Oscillatory flows between deformable boundaries frequently occur in various processes within engineered and biological systems. Oscillatory and, more generally, pulsatile flows enable a broad range of applications, in part because they can emulate physiological flow conditions \cite{DDS20}. However, a complete understanding of \emph{oscillatory} flow in a \emph{compliant two-dimensional (2D) channel} (a canonical configuration) has not been achieved; specifically, the development of the elastoinertial theory and the benchmarking of its predictions against detailed simulation data remain knowledge gaps. 

Classically, oscillatory flows in deformable conduits have been studied in the context of blood vessels in the cardiovascular system and alveoli in the respiratory system, both of which are flexible tubular structures that convey, respectively, blood and air flow \cite{G94,GJ04,HH11}. Oscillatory hydrodynamic forces driven by the heart's pulsations deform blood vessels \cite{P80,F97}. Even beyond the circulatory system, blood flow in the central retinal vein of the eye can be understood as Newtonian fluid flow through a 2D compliant channel in a strongly collapsed state \cite{SF19}. Similarly, in the veins and the peripheral lymphatic system of the brain, flow-induced deformation of compliant structures can provide a valving mechanism, which sets the flow direction \cite{BLCJNB23}. Another example is the synovial joint, a fluid-filled cavity that enables a range of motion in limbs, presenting a 2D micro-elastohydrodynamic lubrication problem \cite{DJ86_2} in which an oscillatory flow couples to a compliant bounding surface \cite{PBG22}. The high lubrication pressures generated during repetitive motion within such narrow confinements can deform the boundaries, as is also the case in roll coating and printing processes \cite{C88,CS97b}. In such problems, also treated as 2D, the interaction between the flow and the deformable surface is significantly influenced by the initial surface topography \cite{YK05}.

In mechanical systems, oscillatory flows between deformable boundaries arise in microfluidic devices fabricated from soft elastomeric materials such as polydimethylsiloxane (PDMS) \cite{XWZZW21,BWW22,Muduetal24}. \emph{Soft hydraulics}, in which internal flow couples with compliant boundary surfaces \cite{C21}, leads to \emph{nonlinear} effects in microfluidic systems, which in turn enable a range of applications, including enhanced fluid mixing \cite{KB16}, flow control \cite{LESKUBL09,Metal10}, and particle manipulation \cite{SD19}, to name a few. Such periodic time-dependent flow can be generated by peristalsis, syringe pumps, as well as other actuation mechanisms driven by electric, magnetic, or acoustic forces \cite{Muduetal24}. Applications of oscillatory flows in engineered microfluidic systems include mimicking physiological processes in organ-on-a-chip technologies \cite{BI14,Lindetal17,DBG22,Leungetal22} and valveless pumping via phase changes and changes in flow resistance \cite{ACB23}. While in some microsystems, coupling of pulsatile flows to elastic boundaries may be beneficial \cite{PKZS14}, it should be avoided in others. This need motivated \citet{BPCOJ22} to develop a bioinspired nonlinear resistor based on soft-hydraulics principles to stabilize the flow rate produced by a peristaltic pump. Deformation of compliant walls due to oscillatory flows is also harnessed in the design of soft robots \cite{Polygerinos17}. For example, \citet{EG14} determined the pressure and deformation profiles in a long, slender cylindrical shell (axisymmetric tube) caused by an oscillatory viscous fluid flow within, as a basic model of a soft robot. The model was extended by \cite{MEG17} to the deformation of slender, beam-like structures filled with a viscous fluid, termed \emph{embedded fluid networks}, leading to novel actuation mechanisms for soft robotic locomotion \cite{MKSGLP23}. Most recently, \citet{NIB25} analyzed, both theoretically and computationally, a 2D soft microfluidic configuration where solute concentration gradients drive elastic deformation of the channel wall, triggering unsteadiness and instability under certain conditions. In a similar vein, \citet{PKMC25,PKMC26} analyzed, both theoretically and computationally, vibrating sheets generating oscillatory flows within thin fluid films in a canonical 2D configuration, determining the conditions under which contactless adhesion or hovering can be achieved. Despite the idealized nature of a 2D configuration, it is relevant to compliant microfluidic systems in which elastomeric microchannels are long and very wide, with one confined, nearly incompressible wall, for which a plane-strain (i.e., 2D) description provides an accurate leading-order model.

The field of \emph{soft lubrication} \cite{SM04} has evolved from the classical field of elastohydrodynamic lubrication \cite{G20}. Specifically, \citet{SM04} identified an optimal combination of geometric and material properties to maximize the normal force exerted by a lubrication flow on an object near a 2D soft wall. The topic of lift forces on objects near elastic boundaries is reviewed in more detail in \cite{KCC18,BCS23,R24}. Subsequently, \citet{CC10} considered the additional effect of a streaming potential (due to electrokinetic phenomena in the flow) on the deformation of the soft layer, while \citet{PKVS16} investigated the case of a viscoelastic 2D soft layer in several canonical configurations. Another variation on the problem, due to \citet{EJG18}, considered a compressible (gas) flow undergoing weak rarefaction through a 2D compliant confinement. However, in most of these preceding works, the displacement of the elastic wall was considered to be primarily in the vertical direction (as justified by the slenderness assumption) and \emph{directly proportional} to the hydrodynamic pressure, which is a Winkler foundation-like mechanism \cite{DMKBF18}. This mechanism is not always the dominant deformation regime, as discussed by \citet{EPKVS21}, who examined the soft lubrication problem for 2D thin and thick elastic layers, the latter \emph{not} being described by a Winkler foundation. A careful asymptotic analysis by \citet{CV20} for 2D slender layers deformed by hydrodynamic forces demonstrated that a nearly incompressible solid yields finite resistance and that the classical Winkler model breaks down. To address disparate results on elastomeric foundations, \citet{CV20} derived a \emph{combined foundation} model that unifies the pressure-deformation relation across the incompressible and compressible regimes of solid deformation, providing a model that is uniformly valid in the limit of an incompressible solid. Importantly, their asymptotic analysis of slender elastic layers also reveals how the transverse components of the displacement depend on the imposed normal load (pressure).

The fluid--structure interaction (FSI) problem, and specifically the two-way pressure-deformation coupling arising from flow in a compliant channel, is central to all the above-mentioned applications. Similar FSI problems also arise in the context of shear-driven flows in 2D confinements with a soft wall driven by the prescribed oscillatory motion of the opposite nonuniform rigid wall \cite{JAS25,TPZR25}. In these problems, flow inertia is neglected, and the \emph{geometric} nonlinearity is treated by a perturbation expansion in the wave amplitude of the rigid wall with prescribed motion. Importantly, in the presence of such nonlinearities, a primary (periodic) flow can give rise to secondary (cycle-averaged, or steady streaming) flow \cite{R01}, referred to as \emph{soft streaming} \cite{BPG22,CBG24}, showing the coupling between \emph{inertial} nonlinearity of the flow and \emph{geometric} nonlinearity due to deformation of a soft structure in an external flow. Meanwhile, despite significant theoretical and computational efforts having been expended on understanding such FSI problems due to internal flows, as discussed above, the \emph{interaction between} geometric and inertial nonlinearities in internal flows evaded attention until the work of \cite{ZR24}. Specifically, they showed that flow inertia provides another nonlinear coupling, tightly linked to deformation, that enhances streaming, a phenomenon termed \emph{elastoinertial rectification}. Previously, \citet{PWC23} studied the two-way coupled oscillatory flow in a 2D rectangular channel and a 3D axisymmetric tube, but neglected flow inertia and did not provide direct numerical simulations for the 2D case. \citet{ZR24} re-analyzed the simulations of \citet{PWC23} for the case of a 3D axisymmetric tube and showed that advective inertia is ``the missing piece'' to fully describe the simulation data. Although \citet{IWC20} previously developed a one-dimensional lubrication model for FSI in a soft-walled microchannel at finite Reynolds number, they considered only the start-up problem and the stability of the steady (inflated) state. Importantly, in following the approach of \cite{ZR24}, we are able to produce accurate reduced-order descriptions of the oscillatory FSI due to internal flow, which are sought after by researchers in biofluid mechanics \cite{CHRTGM06,PV09,QVV16}. As we show, the 2D configuration exhibits qualitatively different physics from its 3D counterparts \cite{ZR24,HPFC25}, including resonance-like behaviors that are absent in those geometries.

The goal of this paper is to address elastoinertial rectification in a 2D fluidic channel with a compliant, nearly incompressible elastic layer forming the top wall. To this end, Sec.~\ref{sec:2D} presents the theoretical formulation and asymptotic analysis of FSI in a 2D channel bounded by a rigid bottom surface and a slender elastic layer above. The analysis involves an oscillatory Newtonian viscous fluid flow interacting with a linearly elastic, nearly incompressible soft solid. First, in Sec.~\ref{sec:gov_eq}, we outline the governing equations for the fluid flow under the lubrication approximation and develop the coupling between fluid and solid mechanics using the combined foundation model proposed by \citet{CV20}. Second, in Sec.~\ref{sec:perturbation}, we perform a perturbation expansion assuming weak compliance, following the methodology of \citet{ZR24}. Using phasors, we derive analytical solutions for both the primary (periodic) and secondary (cycle-averaged, or streaming) flow. Section~\ref{sec:simulations} addresses the computational methodology and direct simulation approach. We provide details of the variational formulation for ALE-FSI, following the framework established by \citet{NZSHK23} and employing quasi-direct coupling between the coupled fluid--solid subproblem and the mesh-motion subproblem. Additionally, we briefly outline the implementation and solution of the variational problem using the FEniCS open-source finite element library. In Sec.~\ref{sec:results}, we present comparisons between theoretical predictions and the computational results, discussing their alignment and providing physical interpretations. Specifically, we validate the theoretical predictions for the pressure distribution, displacement at the fluid--solid interface, and cycle-averaged pressure against ALE-FSI simulations over a range of key dimensionless groups: the Womersley number and the elastoviscous number. Finally, Sec.~\ref{sec:conclusion} concludes the paper. Several appendices provide further details of the analytical calculations in other regimes of interest.


\section{Problem formulation and asymptotic analysis}
\label{sec:2D}

\begin{figure}[ht]
    \centering
    \includegraphics[width=0.65\linewidth]{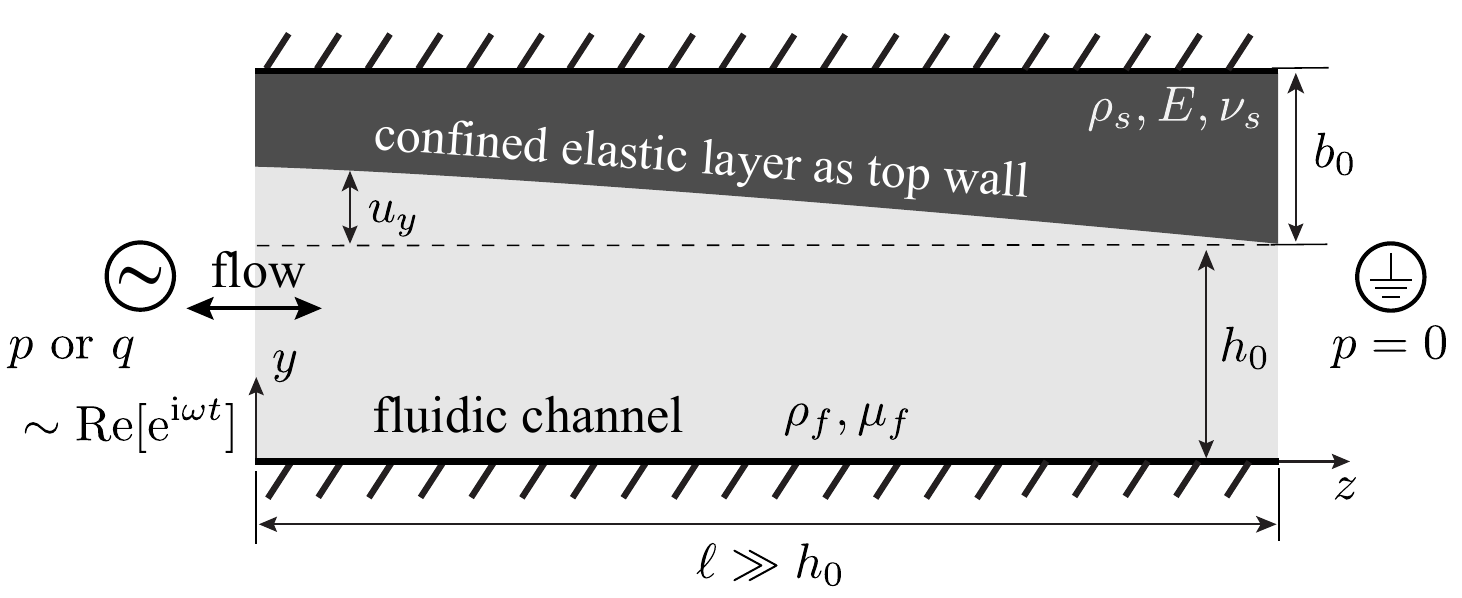}
    \caption{Schematic of the 2D problem of oscillatory flow in a fluidic channel bounded by a rigid surface below and a confined elastic layer above.}
    \label{fig:schematic}
\end{figure}

Consider the oscillatory flow of angular frequency $\omega$ of a Newtonian fluid in a 2D channel in the $(z,y)$ plane as shown in Fig.~\ref{fig:schematic}. The Newtonian fluid has density $\rho_f$ and dynamic viscosity $\mu_f$. The fluidic channel of initial height $h_0$ and length $\ell$ is sandwiched between a rigid surface at $y=0$ and an elastic solid of initial height/thickness $b_0$ and length $\ell$. The elastic solid has density $\rho_s$, shear modulus $G$, and Poisson ratio $\nu_s$. The fluid--solid interface is $y=h_0 + u_y$, where $u_y$ is the vertical displacement into the elastic layer. The solid is clamped such that it has no displacement along the inlet ($z=0$) and outlet ($z=\ell$) planes, ensuring the channel maintains a constant height at its inlet and outlet for all time.

We consider the case of the flow being driven by an oscillatory pressure difference, $\Delta p = p|_{z=0} - p|_{z=\ell} = p_0 \cos(\omega t)$, of magnitude/amplitude $p_0$ and frequency $\omega$, between the inlet and outlet (set to gauge pressure, $p|_{z=\ell}=0$). Then, for a slender channel (such that its  aspect ratio $\epsilon_f={h_0}/{\ell}\ll1$), the suitable nondimensionalization is that arising from lubrication theory:
\begin{equation}
    T = \omega t, \qquad 
    Y = \frac{y}{h_0},\qquad 
    Z = \frac{z}{\ell},\qquad 
    V_Y = \frac{v_y}{\epsilon_f v_c},\qquad 
    V_Z = \frac{v_z}{v_c},\qquad 
    P = \frac{p}{p_c},\qquad 
    U_Y = \frac{u_y}{u_c},\qquad 
    U_Z = \frac{u_z}{u_c},
    \label{eq:ndvars}
\end{equation}
where $v_c = \epsilon_f h_0 p_c/\mu_f$, $u_c = \mathcal{C}_W p_c$, and $p_c$ are velocity, deformation, and pressure scales, respectively. The velocity and pressure scales are related by balancing the $z$-momentum equation of the fluid. The deformation and pressure scales are related by balancing the vertical component of the traction condition at the fluid--solid interface, assuming the solid has a Winkler-like resistance $\mathcal{C}_W^{-1}$ to vertical deformation as its dominant mechanism, that is $u_y \sim \mathcal{C}_W p$ \cite{DMKBF18}. The pressure scale, $p_c=p_0$, can be controlled. Thus, the \emph{compliance number} is $\beta = u_c/h_0 = \mathcal{C}_Wp_0/h_0$, and it quantifies the ``strength'' of the fluid--structure coupling \cite{CCSS17}.

\subsection{Governing equations}
\label{sec:gov_eq}

For a slender conduit, $\epsilon_f\ll1$. Using the variables from Eq.~\eqref{eq:ndvars}, the dimensionless lubrication equations (e.g., \cite{IWC20}) balance flow inertia with pressure gradients in the streamwise $Z$-direction and viscous stresses due to the velocity variation in the perpendicular $Y$-direction:
\begin{subequations}\begin{empheq}[left = \empheqlbrace]{align}
    0 &= - \frac{\partial P}{\partial Y},\label{eq:ymom}\\
    \Wor^2 \left[ \frac{\partial V_Z}{\partial T} + \frac{\beta}{\gamma} \left( V_Y \frac{\partial V_Z}{\partial Y} + V_Z \frac{\partial V_Z}{\partial Z} \right) \right] &= - \frac{\partial P}{\partial Z} + \frac{\partial^2 V_Z}{\partial Y^2}, \label{eq:zmom}\\    
    \frac{\partial V_Y}{\partial Y} + \frac{\partial V_Z}{\partial Z} &= 0. \label{eq:com}
\end{empheq}\label{eq:lubrication_pdes}\end{subequations}
Here, $\Wor=h_0\sqrt{\rho_f \omega/\mu_f}$ is the conventional \emph{Womersley number} \cite{W55a}, while $\gamma = \mathcal{C}_W \ell \mu_f \omega/(\epsilon_f h_0^2)$ is the \emph{elastoviscous number} \cite{ZR24,EG14}. The elastoviscous number is the ratio of timescales---the vertical displacement timescale set by $u_c/(\epsilon_f v_c)$ to the oscillation timescale set by $\omega^{-1}$ \cite{HPFC25}. Table~\ref{table:param_2D_p_control} summarizes the dimensionless numbers of the problem, the assumptions placed on them in the theory, and their representative values considered in the simulations below. These values are generally representative of microfluidic experimental systems \cite{HPFC25}.

\begin{table}
    \centering    
    \begin{ruledtabular}
    \begin{tabular}{l @{\qquad} l @{\qquad} l @{\qquad } l}
        Quantity & Notation & Assumptions in the theory & Simulation value(s) \\
        \hline
        Fluid channel's aspect ratio & $\epsilon_f={h_0}/{\ell}$ & Neglect terms of $\mathrm{O}(\epsilon_f^2)$ & $0.0439$, $0.0358$, $0.0310$, $0.0253$ \\
        Elastic layer's aspect ratio & $\epsilon_s=b_0/\ell$ & Neglect terms of $\mathrm{O}(\epsilon_s^3)$ & $0.0439$, $0.0358$,  $0.0310$, $0.0253$ \\
        Womersley number & $\Wor=h_0\sqrt{\rho_f \omega/\mu_f}$  &Arbitrary& $1,2,3$ \\
        {Elastoviscous number} & $\gamma = \mathcal{C}_W \ell^2 \mu_f \omega/h_0^3$ & Arbitrary& 0.5, 0.75, 1.0, 1.5 \\
        Foundation model's parameter & $\hat{\theta} = \mathcal{C}_I/\mathcal{C}_W$ & Arbitrary & $2.18$ \\
        Foundation model's parameter & $\hat{\vartheta} = \mathcal{C}_H/\mathcal{C}_W$ & Arbitrary & $4$ \\
        Compliance number & $\beta=\mathcal{C}_W p_0/h_0 $ & Neglect terms of $\mathrm{O}(\beta^2)$ &  $0.0034$ \\
        Poisson ratio & $\nu_s$ & Arbitrary in $[0,1/2)$ & $0.45$
    \end{tabular}
    \end{ruledtabular}
    \caption{The dimensionless parameters of the problem of pressure-driven oscillatory flow in a 2D fluidic channel bounded by a slender elastic layer on top under pressure control conditions, i.e., $p_c = p_0$ given. The typical elastic solid is taken to be PDMS \cite{JMTT14}. The fluid properties are varied to yield dimensionless parameters corresponding to the FSI regime of interest. The deformation scale is calculated as $u_c = \beta h_0 = \mathcal{C}_W p_0$ as in \cite{PWC23} and the references therein. The compliances $\mathcal{C}_W = \frac{(1-2\nu_s)}{2(1-\nu_s)}\frac{b_0}{G}$, $\mathcal{C}_I = \frac{2\nu_s(\nu_s-1/4)}{3(1-\nu_s)^2}\frac{b_0}{G}$, and $\mathcal{C}_H = \frac{\nu_s - 1/4}{1 - \nu_s} \frac{b_0}{G}$ are calculated based on \cite{CV20}.}
    \label{table:param_2D_p_control}
\end{table}

The conservation of mass Eq.~\eqref{eq:com} and the kinematic BC, $V_Y|_{Y=H} = (\gamma/\beta) \partial H/\partial T$ (having neglected axial stretching within the lubrication approximation)
are converted into the continuity equation by the usual steps \cite{IWC20}:
\begin{equation}
    \frac{\partial Q}{\partial Z} + \frac{\gamma}{\beta} \frac{\partial H}{\partial T} = 0,
    \label{eq:continuity}
\end{equation}
where the flow rate $q = \int_0^{h_0+u_y} v_z \,dy$ is scaled by $p_0  h_0^3/(\ell \mu_f)$ (per unit width). Notice that the convective term in Eq.~\eqref{eq:zmom} is multiplied by $\beta/\gamma \sim \epsilon_f^2$ but the unsteady term in Eq.~\eqref{eq:continuity} is multiplied by $\gamma/\beta \sim \epsilon_f^{-2}$ (termed a Strouhal number and retained independently of $\epsilon_f$ in some previous works \cite{R83,WW19,IWC20,PWC23}). This observation clarifies that there is a distinguished limit in which $\beta/\gamma$ (or, equivalently, $\gamma/\beta$) is treated as an $O(1)$ term in the lubrication approximation and retained. Importantly, in this distinguished limit, it is possible to linearize the problem while keeping the unsteadiness at the leading order.

For a \emph{confined} and slender elastic layer (as in Fig.~\ref{fig:schematic}), $\epsilon_s \ll 1$. A slender elastic layer's mechanical response to loading is often modeled as a \emph{foundation}, replacing the 2D governing equations of linear elasticity with a relation (i.e., the foundation model) linking the displacement of the fluid--solid interface to the applied pressure (and, possibly, shear) forces \cite{DMKBF18,SM04}. \citet{CV20} developed a combined foundation model that is uniformly valid across the whole range of Poisson ratios (including $\nu_s \to 1/2^-$, i.e., a nearly incompressible yet confined 2D elastic material) by incorporating corrections that become dominant as $\nu_s\to1/2^-$. These higher-order corrections also provide the horizontal component of the fluid--solid interface displacement, which is not available in the leading-order (``Winkler'') foundation model. Adapting their results to the present notation and geometry, we obtain the following expressions for the dimensionless fluid--solid interface displacement components:
\begin{subequations}\begin{align}
    U_Y &= \underbrace{P}_{\text{Winkler}} - \underbrace{\left( \epsilon_s^2 \hat{\theta} \frac{d^2 P}{d Z^2} +\epsilon_f\epsilon_s \hat{\vartheta} \frac{\partial \mathscr{T}_{w}}{\partial Z}\right)}_{\text{corrections to allow $\nu_s\to 1/2^-$}}, \label{eq:combined_foundation_Y}\\
    U_Z &= \underbrace{0}_{\text{Winkler}} - \underbrace{\left(\epsilon_s\frac{h_0}{\mathcal{C}_W G}\mathscr{T}_{w} + \epsilon_s \hat{\vartheta} \frac{d P}{d Z}\right)}_{\text{corrections to allow $\nu_s\to 1/2^-$}},
    \label{eq:combined_foundation_Z}
\end{align}\label{eq:combined_foundation}\end{subequations}
where $\mathscr{T}_{w} \equiv ({\partial V_Z}/{\partial Y})|_{Y=H}$ denotes the shear stress on the elastic wall.
To make Eqs.~\eqref{eq:combined_foundation} dimensionless, we have used the deformation scale $u_c = \mathcal{C}_Wp_0$. Here, $\mathcal{C}_W$ is the compliance introduced by the Winkler ``component'' of the foundation (i.e., $u_y = \mathcal{C}_W p$) \cite{SM04,DMKBF18,BD22}, while $\mathcal{C}_I$ is an analogous compliance for an incompressible foundation (i.e., $u_y = \mathcal{C}_I b_0^2 {d^2 p}/{d z^2}$) \cite{GM70,D89,CV20}. Then, upon making the results of \citet{CV20} dimensionless, we introduce $\theta = \epsilon_s^2 \hat{\theta}$ with $\hat{\theta} = \mathcal{C}_I/\mathcal{C}_W$ and $\vartheta = \epsilon_f\epsilon_s \hat{\vartheta}$ with $\hat{\vartheta} = \mathcal{C}_H/\mathcal{C}_W$, where $\mathcal{C}_H$ is an analogous compliance featured in the horizontal displacement expression. Table~\ref{table:param_2D_p_control} gives representative values; recall that $G$ is the shear modulus, and $\nu_s$ is the Poisson ratio of the material.

In passing, we note the mathematical similarity between Eq.~\eqref{eq:combined_foundation_Y} and the pressure--displacement relationship, $U_Y = P$, for a thin and slender axisymmetric elastic shell \cite{AC18b,ZR24}. The physics are different, however, as this relation arises from hoop stresses in the axisymmetric elastic shell. Finally, an important feature of this 2D confined nearly-incompressible elastic layer is that the shear stresses on the fluid--solid interface scale with $b_0/h_0$ \cite{CV20}, so they cannot be neglected in comparison to pressure gradients if $b_0 \approx h_0$ as in the present work ($\epsilon_s/\epsilon_f=b_0/h_0=1$ in Table~\ref{table:param_2D_p_control}). Furthermore, depending on how close $\nu_s$ is to $1/2$, the factors $\hat{\theta}$, $\hat{\vartheta}$, and $h_0/(\mathcal{C}_W G)$ in Eqs.~\eqref{eq:combined_foundation} can be as large or larger than the aspect ratios premultiplying them \cite{CV20}. As a consequence, the axial displacement $U_Z$, predicted by Eq.~\eqref{eq:combined_foundation_Z}, is not negligible.

\subsection{Perturbation expansion}
\label{sec:perturbation}

Following \citet{ZR24}, we will now seek expansions of all quantities in $\beta\ll1$ (weakly compliant channel) and denote by ``0'' subscripts the ``primary flow'' and by ``1'' subscripts the ``secondary flow.''

First, it is helpful to expand the axial velocity at the deformed interface via a Taylor series expansion in powers of $\beta$ (i.e., domain perturbation; see, e.g., \cite{BPG22,BSC22,ZR24,CBG24}):
\begin{subequations}
\begin{equation}
    V_Z|_{Y=H} = V_{Z,0} |_{Y=1} + \beta \left( V_{Z,1}|_{Y=1} + U_{Y,0} \left.\frac{\partial V_{Z,0}}{\partial Y}\right|_{Y=1}\right) + \mathrm{O}(\beta^2) ,
    \label{eq:VzH_expansion}
\end{equation}
where recall that $H=1+\beta U_Y=1+\beta U_{Y,0} + \mathrm{O}(\beta^2)$.
To enforce the no-slip boundary condition $V_{z}|_{Y=H}=0$, we must then require that
\begin{align}
    V_{Z,0}|_{Y=1} &= 0,\label{eq:Vz0_Y1}\\
    V_{Z,1}|_{Y=1} &= - U_{Y,0} \left.\frac{\partial V_{Z,0}}{\partial Y}\right|_{Y=1}. \label{eq:Vz1_Y1}
\end{align}
Next, the flow rate (volumetric flux) can also be expanded in $\beta$ from its definition in terms of $V_Z$. Using the Leibniz rule to differentiate integrals, and the no-slip condition from Eq.~\eqref{eq:Vz0_Y1}, it is straightforward to show that
\begin{alignat}{2}
    Q_0 &= \lim_{\beta\to0} Q &= \int_0^1 V_{Z,0} \, dY,
    \label{eq:Q0}\\
    Q_1 &= \left.\frac{\partial Q}{\partial \beta} \right|_{\beta=0} &= \int_0^1 V_{Z,1} \, dY.
    \label{eq:Q1}
\end{alignat}%
\end{subequations}

With these results in hand, we break down the coupled problem into solvable $\mathrm{O}(1)$ and $\mathrm{O}(\beta)$ subproblems for the pressure-controlled case. The analogous results for the flow-rate-controlled regime are summarized in Appendix~\ref{app:flow_rate_control}.

\subsubsection{\texorpdfstring{$\mathrm{O}(1)$}{O(1)} solution: Primary flow}
\label{sec:O1_prob}

At $\mathrm{O}(1)$, from Eqs.~\eqref{eq:zmom}, \eqref{eq:continuity}, and \eqref{eq:combined_foundation_Y}, we have:
\begin{subequations}\begin{empheq}[left = \empheqlbrace]{align}
    \Wor^2  \frac{\partial V_{Z,0}}{\partial T} &= - \frac{\partial P_0}{\partial Z} + \frac{\partial^2 V_{Z,0}}{\partial Y^2},\label{eq:Vz0_pde}\\ 
    \theta \frac{d^2 P_0}{d Z^2} + \vartheta \frac{\partial \mathscr{T}_{w,0}}{\partial Z} + U_{Y,0} &= P_0,\label{eq:U0_pde}\\
    \frac{\partial Q_0}{\partial Z} + \gamma \frac{\partial U_{Y,0}}{\partial T} &= 0. \label{eq:Q0_pde}
\end{empheq}\label{eq:O1_prob}\end{subequations}
Compared with previous work \cite{R83,DG91,ZR24,HPFC25}, the key difference here is the form of Eq.~\eqref{eq:U0_pde}, specifically the terms proportional to $\theta$ and $\vartheta$. 
The boundary conditions are no-slip and imposed oscillatory pressure drop, respectively:
\begin{subequations}\begin{empheq}[left = \empheqlbrace]{align}
    V_{Z,0}|_{Y=0} &= V_{Z,0}|_{Y=1} = 0,\\
    P_0|_{Z=0} &= \Real[\re^{\ri T}],\quad P_0|_{Z=1} = 0.
\end{empheq}\label{eq:O1_bc}\end{subequations}

Introducing the phasors
\begin{equation}
    V_{Z,0}(Y,Z,T) = \Real[V_{Z,0,a}(Y,Z)\re^{\ri T}],\qquad Q_{0}(Z,T) = \Real[Q_{0,a}(Z)\re^{\ri T}],\qquad P_{0}(Z,T) = \Real[P_{0,a}(Z)\re^{\ri T}],
\end{equation}
Eqs.~\eqref{eq:O1_prob} and \eqref{eq:O1_bc} become
\begin{subequations}\begin{empheq}[left = \empheqlbrace]{align}
    \Wor^2  \ri V_{Z,0,a} &= - \frac{d P_{0,a}}{d Z} + \frac{d^2 V_{Z,0,a}}{d Y^2}, \label{eq:Vz0_pde_a}\\
    \theta \frac{d^2 P_{0,a}}{d Z^2} + \vartheta \frac{d \mathscr{T}_{w,0,a}}{d Z} + U_{Y,0,a} &= P_{0,a},\label{eq:U0_pde_a}\\
    \frac{d Q_{0,a}}{d Z} + \gamma \ri U_{Y,0,a} &= 0, \label{eq:Q0_pde_a}\\
    V_{Z,0,a}(0) &= V_{Z,0,a}(1) = 0, \label{eq:Vz0_bc_a}\\
    P_{0,a}(0) &= 1,\quad P_{0,a}(1) = 0. \label{eq:P0_bc_a}
\end{empheq}\label{eq:O0_prob_a}\end{subequations}

The complex velocity is easily found from Eqs.~\eqref{eq:Vz0_pde_a} and \eqref{eq:Vz0_bc_a} to be (see also \cite{PWC23}):
\begin{equation}
    V_{Z,0,a}(Y,Z) = 
    \frac{1}{\ri{\Wor}^2}\left[1-\frac{\cos\left(\ri^{3/2} (1-2Y){\Wor}/2\right)}{\cos\left(\ri^{3/2} {\Wor}/2 \right)}\right] \left( -\frac{d P_{0,a}}{d Z} \right),
    \label{eq:Vza0_soln}
\end{equation}
whence
\begin{equation}
    Q_{0,a} = \int_{0}^{1} V_{Z,0,a} \, dY = \underbrace{\frac{1}{\ri \Wor^2}\left[1 - \frac{1}{\ri^{3/2}{\Wor}/2}\tan\left(\ri^{3/2}{\Wor}/2\right)\right]}_{\mathfrak{f}_0(\Wor)} \left(-\frac{d P_{0,a}}{d Z}\right).
    \label{eq:Qa0_soln}
\end{equation}
The dimensionless wall shear stress at the fluid--solid interface in its undeformed position is found from Eq.~\eqref{eq:Vza0_soln} as
\begin{equation}
    \mathscr{T}_{w,0,a} \equiv \left.\frac{\partial V_{Z,0,a}}{\partial Y}\right|_{Y=1} = \underbrace{\frac{\ri^{1/2}}{{\Wor}} \tan\left(\ri^{3/2}{\Wor}/2\right)}_{\mathfrak{f}_1(\Wor)} \left( -\frac{d P_{0,a}}{d Z} \right).
    \label{eq:dVz0dYat1}
\end{equation}

Substituting the flow rate from Eq.~\eqref{eq:Qa0_soln}, the wall shear stress from Eq.~\eqref{eq:dVz0dYat1}, and the displacement solved from Eq.~\eqref{eq:U0_pde_a} into the continuity Eq.~\eqref{eq:Q0_pde_a}, we obtain an ODE for the pressure's phasor amplitude:
\begin{subequations}\begin{empheq}[left = \empheqlbrace]{align}
    \left\{\frac{\mathfrak{f}_0(\Wor)}{\ri \gamma}+\big[\theta-\vartheta \mathfrak{f}_1(\Wor)\big]\right\} \frac{d^2 P_{0,a}}{d Z^2} &=  P_{0,a} , \label{eq:Pa0_ode}\\
    P_{0,a}(0) &= 1,\\
    P_{0,a}(1) &= 0.
\end{empheq}\label{eq:P0a_BVP_pcontrol}\end{subequations}
Furthermore, from Eqs.~\eqref{eq:combined_foundation} and \eqref{eq:Pa0_ode}, we obtain the vertical displacements' phasor amplitudes:
\begin{subequations}\begin{align}
    U_{Y,0,a}(Z) &= \left(\frac{\mathfrak{f}_0}{\ri \gamma}\right) \frac{d^2 P_{0,a}}{d Z^2} = \left(\frac{\mathfrak{f}_0}{\ri \gamma}\right) \kappa^2 P_{0,a}(Z),
    \label{eq:UYa0_soln}\\
    U_{Z,0,a}(Z) &= \left(\epsilon_s\frac{h_0}{\mathcal{C}_W G}\mathfrak{f}_1 -\epsilon_s \hat{\vartheta}\right) \frac{d P_{0,a}}{d Z}.
    \label{eq:UZa0_soln}
\end{align}\label{eq:Ua0_soln}\end{subequations}

The BVP~\eqref{eq:P0a_BVP_pcontrol} is easy to solve, finding
\begin{equation}
    P_{0,a}(Z) = \frac{\sinh\big(\kappa(1-Z)\big)}{\sinh \kappa},\qquad \kappa(\Wor,\gamma,\theta,\vartheta) := \sqrt{\frac{\ri\gamma}{\mathfrak{f}_0(\Wor) + \ri\gamma [\theta-\vartheta \mathfrak{f}_1(\Wor)]}}.
    \label{eq:Pa0_soln}
\end{equation}
Notice that $P_{0,a}(Z) \to (1-Z)$ as $\kappa\to0$ ($\gamma\to0$, or no viscous-elastic coupling, as in a rigid channel).
Variants of this equation were also obtained in \cite{R83,DG91,ZR24} but for 3D axisymmetric tubes and in \cite{HPFC25} for wide 3D channels. While the structure of the solution is similar in all these cases, the form of $\kappa$ differs significantly. As we now show, the complex term, $\ri\gamma [\theta-{\vartheta \mathfrak{f}_1(\Wor)}]$, in the denominator above, leads to significantly different physics of the 2D layer.

\begin{figure}
     \centering
     \begin{subfigure}[b]{0.4\textwidth}
         \centering         
         \includegraphics[width=\textwidth]{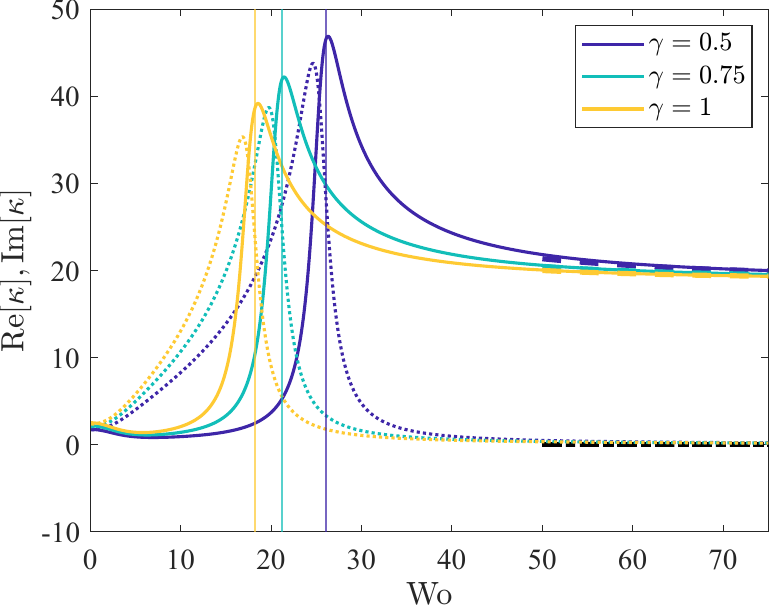}
         \caption{}
     \end{subfigure}
     \qquad\quad
     \begin{subfigure}[b]{0.4\textwidth}
         \centering
         \includegraphics[width=\textwidth]{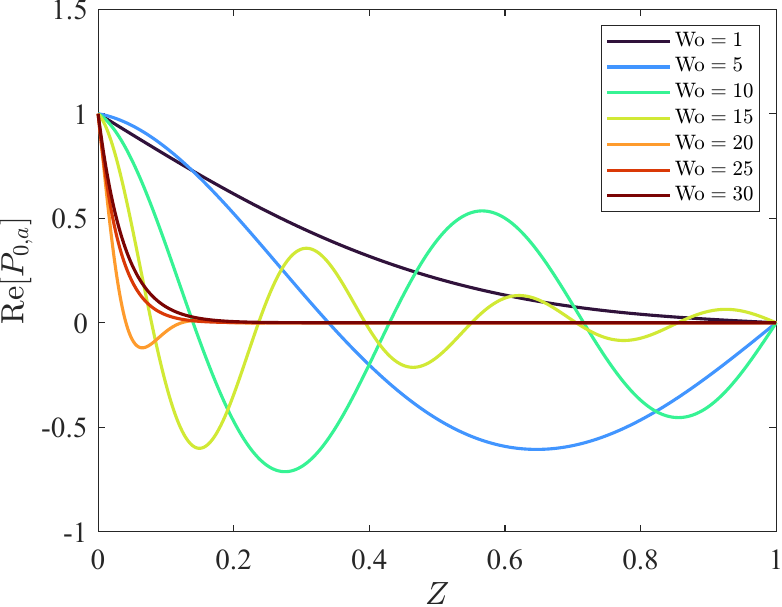}
         \caption{}
     \end{subfigure}
     \caption{(a) The reduced complex ``wavenumber'' $\kappa(\Wor,\gamma,\theta,\vartheta)$ from Eq.~\eqref{eq:Pa0_soln} as a function of $\Wor$, for select values of $\gamma$, and its asymptotic behaviors, for $\theta=0.0028$ and $\vartheta=0.0051$. The solid and dotted curves represent, respectively, $\Real[\kappa]$ and $\Imag[\kappa]$, and the dashed curves represent the large-$\Wor$ asymptotics, shown with the same color. The bold dash-dotted line on the right is the axis, shown again for clarity. The thin vertical lines show the approximation to the Womersley number [$\Wor_c$ from Eq.~\eqref{eq:Wor_c}] at which the peak occurs. (b) Shape of the primary pressure amplitude's axial distribution, $\Real[P_{0,a}(Z)]$ from \eqref{eq:Pa0_soln}, for $\gamma = 0.75$,  $\theta=0.0028$ and $\vartheta=0.0051$, across a range of $\Wor$ numbers, showing the effect of the peak in (a).}
     \label{fig:kappa_effects}
\end{figure}

As noted by \citet{ZR24}, $P_{0,a}(Z)$ ``does not vary linearly along the length of a deformable channel, but instead decays exponentially with spatial oscillations,'' characterized by the complex ``wavenumber'' $\kappa$ in Eq.~\eqref{eq:Pa0_soln}. Notice, however, that the combined foundation model leads $\kappa$ to depend on $\gamma$ in a nonrescalable way (unlike in \cite{ZR24} or \cite{HPFC25}, wherein $\kappa/\sqrt{\gamma}$ is solely a function of $\Wor$). 
Figure~\ref{fig:kappa_effects}(a) shows the plot of the complex ``wavenumber'' $\kappa$ variation with $\Wor$ at three different values of $\gamma$. We also show the small- and large-$\Wor$ asymptotics, calculated as:
\begin{equation}
    \kappa(\Wor,\gamma,\theta,\vartheta) \sim 
    \begin{cases}
    2 \sqrt{\frac{3\ri \gamma}{6\ri \gamma(2\theta+\vartheta) + 1}} \left\{ 1 + \frac{1+5\ri\gamma\vartheta}{20} \left[\frac{\ri\gamma}{6 \ri \gamma(2\theta+\vartheta)  + 1}\right] {\Wor}^2 \right\}, &\quad  \Wor\rightarrow 0,\\[10pt]
    \begin{cases}\frac{1}{\sqrt{\theta}} \left( 1 + \frac{\ri^{3/2}\vartheta}{2\theta \Wor}+\frac{4\theta-3\ri\gamma\vartheta^2}{8\gamma\theta^2\Wor^2}\right), &\quad \theta,\vartheta \ne 0,\\[10pt]
    \sqrt{\gamma} \left(\ri \Wor + \sqrt{\ri} + \frac{3}{2\Wor}\right), &\quad \theta = \vartheta = 0,
    \end{cases}
    &\quad \Wor\rightarrow\infty.
    \end{cases}
\end{equation}

Comparing these asymptotics to those obtained by \citet{ZR24} and \citet{HPFC25}, we observe that in the limit of  $\Wor\rightarrow \infty$, there are two different limiting behaviors for $\kappa$, depending on the combined foundation's parameters. When $\theta=\vartheta=0$ (pure Winkler foundation), we recover the asymptotics of \citet{HPFC25} for a 3D rectangular configuration. When $\theta,\vartheta\neq0$, however, the large-$\Wor$ asymptotics are controlled by $1/\sqrt{\theta}$, independently of $\Wor$. This novel behavior is due to the mechanics of the confined, nearly incompressible 2D elastic layer that bounds the fluidic channel, specifically the nonlocal pressure--displacement coupling arising in this configuration, which is properly built into the combined foundation model. This regime also allows for the possibility of extrema with respect to $\Wor$, since $\kappa$ saturates as $\Wor\to\infty$.

Indeed, we observe peaks (maxima of $\kappa$) in Fig.~\ref{fig:kappa_effects}(a) for specific values of $\Wor$, at each value of $\gamma$. This behavior was not observed in prior work \cite{ZR24,HPFC25}. Following \citet{YCGR20}, we find an approximation for this value $\Wor_c$ by balancing the smallest terms in the denominator of $\Imag[\kappa]$ for $\Wor\gg1$. The value $\Wor_c$ approximately minimizes the denominator, thus approximately maximizes $\Imag[\kappa]$. We find that
\begin{equation}
    \Wor_c \approx -\dfrac{1}{2\sqrt{2}} \left(\dfrac{\vartheta}{\theta}\right) + \sqrt{ \frac{1}{8}\left(\dfrac{\vartheta}{\theta}\right)^2 + \frac{1}{\theta\gamma}}
    \label{eq:Wor_c}
\end{equation}
is within $5\%$ of the $\Wor$ values at which the peaks occur, for all choices of $\gamma$ shown.

To highlight the effect of these peaks, and the meaning of $\Wor_c$, we plot the amplitude of the pressure distribution $\Real[P_{0,a}(Z)]$, in Fig.~\ref{fig:kappa_effects}(b). Near  $\Wor = \Wor_c \approx 20$ (for $\gamma=0.75$), we see the oscillatory nature of the pressure distribution, which is in contrast with its rapidly decaying nature for values of $\Wor$ much above $\Wor_c$. This behavior suggests a ``cutoff'' frequency (equivalently, ``cutoff'' value of $\Wor$) in the system, which could be exploited for design purposes. The cutoff can be understood by noting that both $\Real[\kappa]$ and $\Imag[\kappa]$ peak near this value of $\Wor$, suggesting similar rates of spatial decay and oscillation. Then, in complete contrast to the previous works \cite{ZR24,HPFC25}, $\Imag[\kappa]$ rapidly decays to zero for large $\Wor$, rendering the pressure profile mostly monotone with a rapid spatial decay rate of $\propto 1/\sqrt{\theta}$.

\subsubsection{\texorpdfstring{$\mathrm{O}(\beta)$}{O(beta)} solution: Secondary flow}
\label{sec:Obeta_prob}

At $\mathrm{O}(\beta)$, from Eqs.~\eqref{eq:zmom} and \eqref{eq:continuity}, we have:
\begin{subequations}\begin{empheq}[left = \empheqlbrace]{align}
    \Wor^2 \left[ \frac{\partial V_{Z,1}}{\partial T} + \frac{1}{\gamma} \left( V_{Y,0} \frac{\partial V_{Z,0}}{\partial Y} + V_{Z,0} \frac{\partial V_{Z,0}}{\partial Z} \right) \right] &= - \frac{\partial P_1}{\partial Z} + \frac{\partial^2 V_{Z,1}}{\partial Y^2}, \label{eq:Vz1_pde}\\
    \frac{\partial Q_1}{\partial Z} + \gamma \frac{\partial U_{Y,1}}{\partial T} &= 0. \label{eq:Q1_pde}
\end{empheq}\label{eq:O2_prob}\end{subequations}
And, using Eq.~\eqref{eq:VzH_expansion}, the boundary conditions are:
\begin{subequations}\begin{empheq}[left = \empheqlbrace]{alignat=2}
    V_{Z,1}|_{Y=0} &= 0,\qquad & V_{Z,1}|_{Y=1} &= - U_{Y,0} (\partial V_{Z,0}/\partial Y)|_{Y=1},\\
    P_1|_{Z=0} &= 0,\qquad & P_1|_{Z=1} &= 0.
\end{empheq}\label{eq:O2_bc}\end{subequations}
Recall that, for two phasors $A = A_a \re^{\ri T}$ and $B = B_a \re^{\ri T}$, $\langle A B \rangle = \tfrac{1}{2}\Real[A_a^*B_a] = \tfrac{1}{2}\Real[A_aB_a^*]$, where the star superscript denotes the complex conjugate. In Eq.~\eqref{eq:Vz1_pde}, $V_{Y,0}$ is to be determined from $V_{Z,0}$ via Eq.~\eqref{eq:com} at $\mathrm{O}(\beta^0)$ and the no-penetration condition at $Y=0$.

Being interested in the streaming/rectified flow, following \citet{ZR24}, we cycle-average the $\mathrm{O}(\beta)$ problem using $\langle \, \cdot \, \rangle := \frac{1}{2\pi}\int_0^{2\pi} (\,\cdot\,) \,dT$, understanding that the underlying quantities are time-periodic, so that $\langle \partial (\,\cdot \,)/\partial T \rangle = 0$. We obtain
\begin{subequations}\begin{empheq}[left = \empheqlbrace]{align}
    \frac{\Wor^2}{\gamma} \left\langle V_{Y,0} \frac{\partial V_{Z,0}}{\partial Y} + V_{Z,0} \frac{\partial V_{Z,0}}{\partial Z} \right\rangle &= - \frac{d \langle P_1\rangle }{d Z} + \frac{\partial^2 \langle V_{Z,1} \rangle}{\partial Y^2}, \label{eq:Vz1_avg_pde}\\
    \frac{\partial \langle Q_1 \rangle}{\partial Z}  &= 0. \label{eq:Q1_avg_pde}
\end{empheq}\label{eq:O2_avg_prob}\end{subequations}
The boundary conditions are:
\begin{subequations}\begin{empheq}[left = \empheqlbrace]{alignat=2}
    \langle V_{Z,1} \rangle|_{Y=0} &= 0,\qquad & \langle V_{Z,1} \rangle |_{Y=1} &= - \langle U_{Y,0} (\partial V_{Z,0}/\partial Y) |_{Y=1} \rangle, \label{eq:Vz1_avg_bc}\\
    \langle P_1 \rangle|_{Z=0} &= 0,\qquad & \langle P_1 \rangle|_{Z=1} &= 0. \label{eq:P1_avg_bc}
\end{empheq}\label{eq:O2_avg_bc}\end{subequations}

As in \cite{ZR24}, we observe that Eq.~\eqref{eq:Vz1_avg_pde} has a solution of the form
\begin{equation}
    \langle V_{Z,1} \rangle = - \frac{1}{2}\frac{d \langle P_1\rangle }{d Z} Y(1-Y) - Y \left.\left\langle U_{Y,0} \frac{\partial V_{Z,0}}{\partial Y} \right|_{Y=1}\right\rangle + \widetilde{\langle V_{Z,1} \rangle},
    \label{eq:Vz1T}
\end{equation}
which satisfies the BCs~\eqref{eq:Vz1_avg_bc}, and reduces Eq.~\eqref{eq:Vz1_avg_pde} to
\begin{equation}
    \frac{\Wor^2}{\gamma} \left\langle V_{Y,0} \frac{\partial V_{Z,0}}{\partial Y} + V_{Z,0} \frac{\partial V_{Z,0}}{\partial Z} \right\rangle =  \frac{\partial^2 \widetilde{\langle V_{Z,1} \rangle}}{\partial Y^2},\qquad \widetilde{\langle V_{Z,1} \rangle}|_{Y=0} = \widetilde{\langle V_{Z,1} \rangle}|_{Y=1} = 0.
    \label{eq:Vz1T_avg_pde}
\end{equation}

While the left-hand side of Eq.~\eqref{eq:Vz1T_avg_pde} \emph{can} be evaluated as an explicit function of $Y$, it seems challenging to solve the ODE for $\widetilde{\langle V_{Z,1} \rangle}$. At any rate, it can be evaluated numerically by solving Eq.~\eqref{eq:Vz1T_avg_pde} as a two-point BVP for $\widetilde{\langle V_{Z,1} \rangle}(Y)$. We do so using both the symbolic and numeric computing capabilities of \textsc{Matlab} to form the LHS exactly, then solve the BVP using \texttt{bvp4c} (absolute and relative tolerance of $10^{-6}$) \cite{KS01}.

Finally, from Eqs.~\eqref{eq:Q1} and \eqref{eq:Vz1T}, we have
\begin{equation}
    \langle Q_1 \rangle = \int_0^1 \langle V_{Z,1} \rangle \, dY   
    = -\frac{1}{12}\frac{d \langle P_1\rangle }{d Z} - \frac{1}{2}\left.\left\langle U_{Y,0} \frac{\partial V_{Z,0}}{\partial Y} \right|_{Y=1}\right\rangle + \int_0^1 \widetilde{\langle V_{Z,1} \rangle} \, dY.
\end{equation}
Noting that $\langle Q_1 \rangle$ is a constant per Eq.~\eqref{eq:Q1_avg_pde}, the last equation can be integrated in $Z$. Then, using the BC~\eqref{eq:P1_avg_bc}, we have
\begin{equation}
    \langle Q_1 \rangle = \int_0^1 \left[-\frac{1}{2}\left.\left\langle U_{Y,0} \frac{\partial V_{Z,0}}{\partial Y} \right|_{Y=1}\right\rangle +  \int_0^1 \widetilde{\langle V_{Z,1} \rangle} \, dY \right]dZ,
    \label{eq:Q1_avg}
\end{equation}
and
\begin{equation}
    \langle P_1 \rangle(Z) = 12 \int_0^Z \left[  -\frac{1}{2}\left.\left\langle U_{Y,0} \frac{\partial V_{Z,0}}{\partial Y} \right|_{Y=1}\right\rangle + \int_0^1 \widetilde{\langle V_{Z,1} \rangle} \, dY \right] dZ - 12 Z \langle Q_1 \rangle.
    \label{eq:P1_avg}
\end{equation}

Finally, we emphasize (as shown in \cite{ZR24} for the axisymmetric case), that neglecting the left-hand side of Eq.~\eqref{eq:Vz1T_avg_pde}, and thus the second term in the integration in Eq.~\eqref{eq:P1_avg} (as was done in \cite{PWC23}), leads to uncontrolled errors of $\mathrm{O}(1)$ in $\langle P_1 \rangle$.

\begin{figure}
     \centering
     \begin{subfigure}[b]{0.4\textwidth}
         \centering         
         \includegraphics[width=\textwidth]{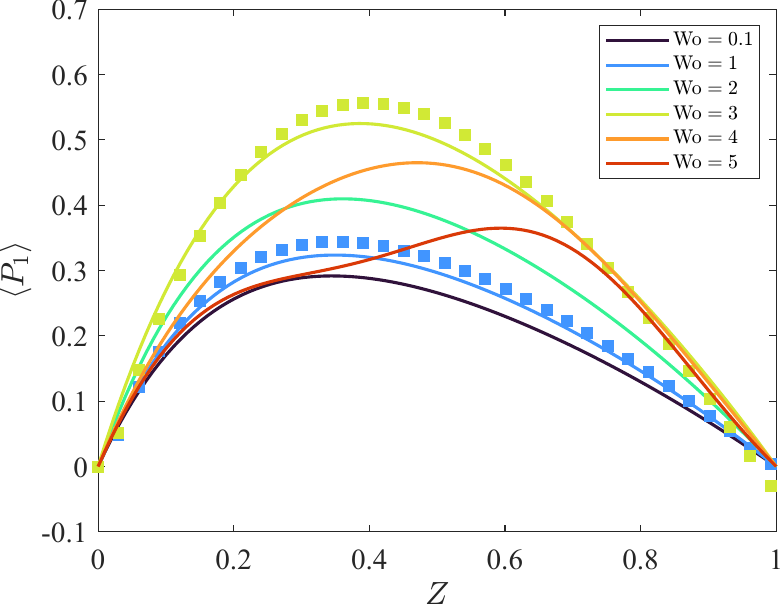}
         \caption{}
     \end{subfigure}
     \qquad\quad
     \begin{subfigure}[b]{0.4\textwidth}
         \centering
         \includegraphics[width=\textwidth]{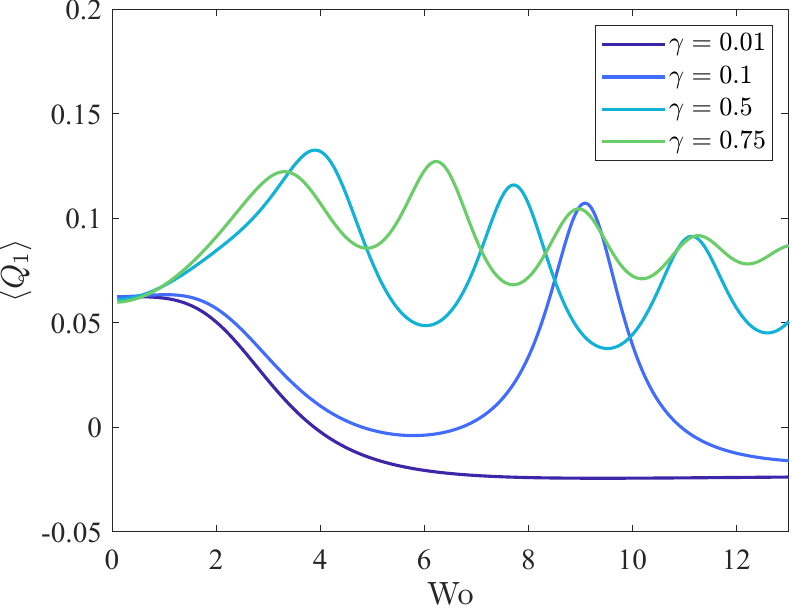}
         \caption{}
     \end{subfigure}
     \caption{(a) The streaming pressure profile from Eq.~\eqref{eq:P1_avg} for $\gamma=0.75$, highlighting the resonance-like behavior for $2 \lesssim \Wor \lesssim 5$. (b) The streaming flux $\langle Q_1\rangle$ from Eq.~\eqref{eq:Q1_avg}, as functions of $\Wor$. Both panels are for $\theta=0.0028$ and $\vartheta=0.0051$. Symbols in (a) correspond to finite element simulations described in Sec.~\ref{sec:simulations}.}
     \label{fig:theoretical time-averaged-flux}
\end{figure}

To illustrate these results, Fig.~\ref{fig:theoretical time-averaged-flux}(a) shows the streaming pressure profile $\langle P_1 \rangle(Z)$ at a fixed value of $\gamma=0.75$ and various values of $\Wor$ to highlight the nonmonotonic dependence on $\Wor$. Meanwhile, Fig.~\ref{fig:theoretical time-averaged-flux}(b) shows the variation of the streaming flux $\langle Q_1 \rangle$ with $\Wor$ at four different values of $\gamma$, showing that as $\gamma$ increases, $\langle Q_1 \rangle$ is also a nonmonotonic function of $\Wor$, reminiscent of ``resonances'' that extremize the streaming flux. Focusing on the case of $\gamma = 0.75$, Fig.~\ref{fig:theoretical time-averaged-flux}(a) shows that there is a ``sudden'' large enhancement in the maximum values of $\langle P_1\rangle$ near $\Wor \approx 3$. Likewise, we observe that $\langle Q_1\rangle$ in Fig.~\ref{fig:theoretical time-averaged-flux}(b) increases by an order of magnitude as well near $\Wor \approx 3$. Interestingly, the flow rate enhancement is oscillatory in Fig.~\ref{fig:theoretical time-averaged-flux}(b) and can change sign with $\gamma$, as also discussed in \cite{ZR24}. Finally, the calculations in Appendix~\ref{app:approx_W0} show that $\langle Q_1 \rangle \to 1/16 = 0.0625$ minus an $\mathcal{O}(\gamma^2)$ correction as $\Wor \to 0$, consistent with Fig.~\ref{fig:theoretical time-averaged-flux}(b).

For larger values of $\gamma$, Fig.~\ref{fig:theoretical time-averaged-flux}(b) shows that the number of peaks in $\langle Q_1 \rangle$ increases, suggesting now a sequence of unevenly-spaced resonances. This behavior is challenging to explain intuitively as effective slip at the top wall's unperturbed location (due to fluid--structure interaction) and the flow's advective inertia, which set the behavior of $\langle Q_1 \rangle$, are nonlinearly coupled. Yet, there are apparently ``trade-offs'' between the two effects as the primary flow's structure changes with $\Wor$, leading to specific resonant frequencies (equivalently, $\Wor$ values). This behavior is reminiscent of scalar transport in oscillatory flows of viscoelastic fluids, in which such resonances also occur in the effective dispersion or effective flux of the passive scalar \cite{YCGR20,MABM25}.

These observations about how $\langle P_1 \rangle$ depends on $\Wor$ are to be contrasted with the 3D rectangular \cite{HPFC25} and 3D axisymmetric \cite{ZR24} geometries, in which $\langle P_1 \rangle$ does not exhibit any nodes (change of sign of $\langle P_1 \rangle$ with $Z$) as $\Wor$ is varied. Furthermore, even though the magnitude of $\langle P_1 \rangle$ does show some weak nonmonotonic dependence with $\Wor$ for the case of a 3D axisymmetric thin shell \cite{ZR24}, it is not as pronounced as in Fig.~\ref{fig:theoretical time-averaged-flux}(a), in which there is up to a factor of $2.5$ increase in the magnitude as $\Wor$ goes from $0.1$ to $3$. These observations suggest an intricate and strong coupling between oscillatory flow and deformation of a confined nearly-incompressible elastic layer. The key point is that the deformation of this elastic layer, modeled as a combined foundation here ($\theta, \vartheta \ne 0$), represents markedly different physics than the unconfined cases \cite{HPFC25,ZR24}, which are well described by a Winkler-foundation-type mechanism ($\theta = \vartheta = 0$).


\section{Computational model and simulation approach}
\label{sec:simulations}

\begin{figure}[ht]
    \centering
    \includegraphics[width=0.5\textwidth]{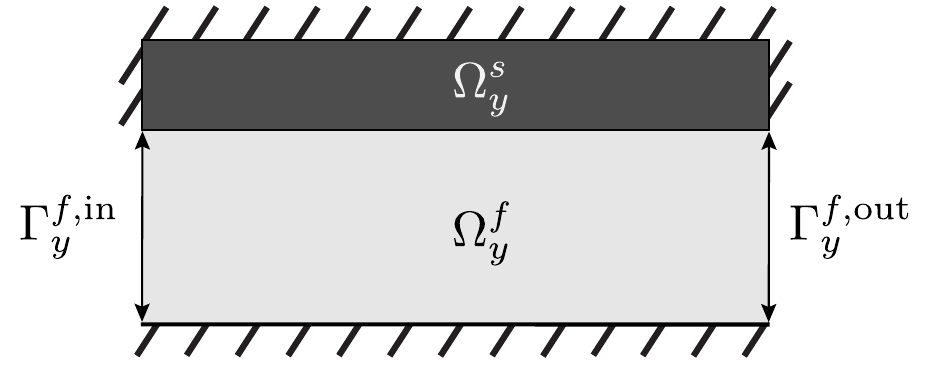}
    \caption{The 2D {referential} computational domain comprising distinct fluid and solid subdomains used to simulate the flow and deformation shown schematically in Fig.~\ref{fig:schematic}.}
    \label{fig:2DDomain}
\end{figure}

Figure~\ref{fig:2DDomain} shows the schematic of the 2D \emph{referential} domain $\Omega_y \subset \mathbb{R}^2$ for the computational model. The referential domain has two subdomains, the fluid subdomain $\Omega_y^{f}$ and the solid subdomain $\Omega_y^{s}$, such that $\Omega_y = \overline{\Omega_y^{f} \cup \Omega_y^{s}}$. Since the two subdomains are nonoverlapping, $\Omega_y^{f} \cap \Omega_y^{s} = \emptyset$. Analogously, we can define the \emph{spatial} domain $\Omega_x \subset \mathbb{R}^2$.

In this model of the flow in a 2D deformable channel,  $\Omega_y^{s}$ is the channel wall, which is considered a hyperelastic solid material that undergoes mild deformation because of the hydrodynamic force applied by the Newtonian incompressible fluid flow within $\Omega_y^f$. Consistent with Sec.~\ref{sec:2D}, the elastic wall is constrained on its left, right, and top edges to have zero displacement (clamped condition). In the fluid domain $\Omega_y^f$, the bottom wall is rigid, and the fluid flow is subjected to the no-slip and no-penetration boundary conditions. The fluid enters through the inlet of the channel $\Gamma_y^{f,\mathrm{in}}$. In the pressure-controlled case, a uniform oscillatory pressure is applied as a Neumann boundary condition. The fluid exits the channel through the outlet $\Gamma_y^{f,\mathrm{out}}$, where the traction is set to zero as an outflow boundary condition. 

\subsection{The arbitrary Lagrangian--Eulerian description and the weak form}

To simulate the problem, we used an arbitrary Lagrangian--Eulerian (ALE) description for the fluid subproblem and a Lagrangian description for the solid subproblem. We follow the kinematic framework established by \citet{BCHZ08}, and the notation used follows \cite{NZSHK23,BCHZ08}. 
The ALE description of the fluid subproblem takes the form
\begin{subequations}\begin{empheq}[left = \empheqlbrace]{align}
    \rho_f \left(\frac{\partial \bm{v}}{\partial t} + (\bm{v}-\hat{\bm{v}}) \cdot \nabla_{\bm{x}} \bm{v}\right) - \nabla_{\bm{x}} \cdot \boldsymbol{\sigma}_f &= \bm{0},
    \label{eq:lin mom fluid}\\
    \nabla_{\bm{x}} \cdot \bm{v} &= 0, 
    \label{eq:mass fluid}
\end{empheq}\label{eq:fluid}\end{subequations}
where $\bm{\sigma}_f = 2\mu_f \symm(\nabla_{\bm{x}} \bm{v}) - p\mathbf{I}$ is the stress tensor and $\symm(\nabla_{\bm{x}} \bm{v})$ is the symmetric part of the gradient tensor. 
Equation~\eqref{eq:lin mom fluid} represents the balance of linear momentum for the incompressible Newtonian fluid subproblem, and Eq.~\eqref{eq:mass fluid} represents mass conservation for the incompressible fluid. 

In addition to the Neumann (pressure) boundary condition at the inlet and the zero traction boundary condition at the outlet mentioned above, we further impose the continuity of traction at the fluid--solid subdomains' interface. The Neumann boundary conditions are inserted directly into the weak form of the problem. Then, following \cite{NZSHK23}, the residual of the weak formulation for Eqs.~\eqref{eq:fluid} and their BCs on the deforming domain is given by
\begin{equation}
    \begin{aligned}
    R_f\left(\left\{\bm{v}, p \right\}, \left\{\bm{w}, q \right\}\right) 
    &= \int_{\Omega^f_x} \rho_f 
    \left( \left. \frac{\partial \bm{v}}{\partial t} \right|_{\bm{y}} 
    + (\bm{v} - \hat{\bm{v}}) \cdot \nabla_{\bm{x}} \bm{v} \right) 
    \cdot \bm{w} \, d\Omega_x \\
    &\quad + \int_{\Omega^f_x} \boldsymbol{\sigma}_f : \nabla_{\bm{x}} \bm{w} \, d\Omega_x  + \int_{\Gamma_{x}^{f,\mathrm{in}}} p_\mathrm{in} \bm{n}_x \cdot \bm{w} \, d\Gamma_x  \\
    &\quad + \int_{\Omega^f_x} q \nabla_{\bm{x}} \cdot \bm{v} \, d\Omega_x  \\
    &\quad - \int_{\Gamma_{x}^{f,\mathrm{out}}} 
    \rho_f \left(\left\{(\bm{v} - \hat{\bm{v}}) \cdot \bm{n}_x\right\}_- \bm{v}\right) 
    \cdot \bm{w} \, d\Gamma_x,
    \end{aligned}
    \label{eq:weak fluid}
\end{equation}
where $\bm{w}$, $q$ (not to be confused with flow rate) are suitable test functions, discussed below, and the notation $\frac{\partial(\,\cdot\,)}{\partial t}|_{\bm{y}}$ emphasizes that this time derivative is on the referential domain. The last term of $R_f$ in Eq.~\eqref{eq:weak fluid} is for backflow stabilization at the outlet \cite{BGHMZ09,EBHVM11}, where
\begin{equation}
    \left\{(\bm{v} - \hat{\bm{v}}) \cdot \bm{n}_x \right\}_- = \frac{1}{2}\left\{(\bm{v} - \hat{\bm{v}}) \cdot \bm{n}_x - \lvert (\bm{v} - \hat{\bm{v}}) \cdot \bm{n}_x \rvert \right\}.
\end{equation}
In Eq.~\eqref{eq:weak fluid}, the inlet boundary conditions dictate that 
\begin{subequations}\begin{align}
    p_\mathrm{in} &=p_0 \cos(\omega t),\\
    \hat{\bm{v}} &= \left. \frac{\partial \hat{\bm{u}}}{\partial t} \right|_{\bm{y}},
\end{align}\end{subequations}
where $\hat{\bm{u}}$ is the referential displacement field, that is $\bm{x}=\bm{y}+\hat{\bm{u}}$.

For the solid subproblem, the balance of linear momentum in the Lagrangian frame is
\begin{equation}
    \rho_{s} \frac{\partial^2 \bm{u}_{s}}{\partial t^2} - \nabla_{\bm{X}} \cdot (\mathbf{FS}) = \bm{0},
    \label{eq:pde_us}
\end{equation}
where $\bm{u}_{s}$ is the solid's displacement field, $\mathbf{F} = \mathbf{I} + \nabla_{\bm{X}} \bm{u}_{s}$ is the deformation gradient associated with the mapping of the material domain to its spatial domain, and  $\mathbf{S}$ is the second Piola--Kirchhoff stress tensor. In Eq.~\eqref{eq:pde_us}, it is understood that $\rho_s$ is the mass density of the solid in the initial configuration, and the ``material'' and ``referential'' clocks are synchronized  \cite{BCHZ08}.  
As in \cite{NZSHK23}, the velocity field can be expressed as the time derivative of a displacement field as
\begin{subequations}\label{eq:ale_fsi_coupling}\begin{align}
     \bm{v}_{fs}  &= \left. \frac{\partial \bm{u}_{s}}{\partial t} \right|_{\bm{X}}, \label{eq:ale_fsi_coupling_v}\\
    \hat{\bm{u}} &= \bm{u}_{s}, \label{eq:ale_fsi_coupling_u}
\end{align}\end{subequations}
in $\Omega^{s}_y$. Together, these equations enforce the two kinematic interface conditions required for two-way coupling in the ALE-FSI formulation: Eq.~\eqref{eq:ale_fsi_coupling_v} ensures continuity of velocity at the fluid--solid interface (no-slip and no-penetration), while Eq.~\eqref{eq:ale_fsi_coupling_u} ties the ALE mesh displacement to the solid displacement, so that the fluid mesh conforms to the deforming interface at all times.

Based on Eq.~\eqref{eq:pde_us}, the residual of the weak formulation for the solid subproblem is given by
\begin{equation}
    R_{s}\left(\left\{\bm{v}, \bm{w} \right\}\right) = \int_{\Omega^{s}_y} \rho_s \left. \frac{\partial \bm{v}}{\partial t} \right|_{\bm{X}} \cdot \bm{w} + (\mathbf{FS}): \nabla_{\bm{X}} \bm{w} \ d \Omega_y.
    \label{eq:weak solid}
\end{equation}
We employed the neo-Hookean hyperelastic model with an isochoric--volumetric split \cite[Ch.~6]{H00}, for which the second Piola--Kirchhoff stress is
\begin{equation}
    \mathbf{S} = \underbrace{\mu_s J^{-2/3} \left( \mathbf{I} - \tfrac{1}{3} I_1\mathbf{C}^{-1} \right)}_{\text{isochoric}} + \underbrace{KJ(J-1)\mathbf{C}^{-1}}_{\text{volumetric}} \, ,
    \label{eq:2nd piola stress}
\end{equation}
where $J = \det\mathbf{F}$, $\mathbf{C} = \mathbf{F}^\top \mathbf{F}$ is the right Cauchy--Green tensor, $I_1$ is its first invariant, $K = \lambda_s + 2\mu_s/3$ is the bulk modulus, and $\lambda_s=2G\nu_s/(1-2\nu_s)$, $\mu_s=G$ are the Lam\'e constants. Note that many different volumetric strain energy functions have been proposed \cite{H00,DS00b}, yielding slightly different versions of the second term in Eq.~\eqref{eq:2nd piola stress}; we have used the simplest one \cite{DS00b,AMSS25}. 
For the 2D configuration at hand, the ALE-FSI problem is solved in a plane strain configuration, for which $\mathbf{C}$ is $2\times2$, we calculate $I_1 = \tr(\mathbf{C}) + 1$; the $+1$ represents the unit out-of-plane stretch and must be retained to avoid artificially softening the material \cite{AMSS25}. Importantly, the neo-Hookean model reduces to linear elasticity with the same Lam\'{e} constants for small strains, so the comparison to the theory is asymptotically justified for $\beta \ll 1$.

The complete fluid--solid problem residual then becomes
\begin{equation}
     R_{fs}\left(\left\{\bm{v}, p \right\}, \left\{\bm{w}, q \right\}\right) =  R_f\left(\left\{\bm{v}, p \right\}, \left\{\bm{w}, q \right\}\right) + R_{s}\left(\left\{\bm{v}, \bm{w} \right\}\right) + \text{stabilization terms}.
     \label{eq:fluid-solid residual}
\end{equation}
In Eq.~\eqref{eq:fluid-solid residual}, SUPG/PSPG and LSIC stabilization terms are added based on the fluid's momentum equation and mass conservation equation, respectively. A detailed description of these stabilization terms can be found in the textbook of \citet{BTT13}. In the remainder of this work, whenever we refer to Eq.~\eqref{eq:fluid-solid residual}, it is implied that the SUPG/PSPG and LSIC stabilization terms are included, but we will not write them out for brevity. 

In passing, we remind the reader that the continuity of traction along the fluid--solid interface is automatically satisfied when using a combined residual $R_{fs}$ as in Eq.~\eqref{eq:fluid-solid residual} along with the integrated-by-parts fluid and solid residuals, $R_f$ in Eq.~\eqref{eq:weak fluid} and $R_s$ in Eq.~\eqref{eq:weak solid}. Specifically, the interface traction terms are equal and opposite, canceling in $R_{fs}$, so we have not written them in Eqs.~\eqref{eq:weak fluid} and \eqref{eq:weak solid} above.

Now that we have specified the net residual for the coupled fluid--solid subproblem, we define the mesh motion subproblem, in which we solve for $\hat{\bm{u}}$. We treat the mesh motion subproblem as a fictitious elastic-like problem with $\hat{\mathbf{S}} = \hat{K} \tr(\hat{\mathbf{E}}) \mathbf{I} + 2 \hat{\mu} \big[\hat{\mathbf{E}} - \tfrac{1}{3}\tr(\hat{\mathbf{E}})\mathbf{I}\big]$, where $\hat{\mathbf{E}} = \tfrac{1}{2}(\hat{\mathbf{F}}^\top\hat{\mathbf{F}} - \mathbf{I})$ is the Green--Lagrange strain tensor based on the mesh deformation tensor $\hat{\mathbf{F}} = \mathbf{I} + \nabla_{\bm{X}} \hat{\bm{u}}$. The artificial material parameters are chosen as $\hat{K} = \hat{\mu} = (\det\hat{\mathbf{F}})^{-3}$ to implement Jacobian-based mesh stiffening (see, e.g., \cite{STB03,BTT13,NZSHK23} and the references therein).

\subsection{Implementation in FEniCS}
We posed all residuals (fluid, solid, and mesh) back to the static reference domain, on which the function spaces for the finite element method were constructed. The method to solve these variational forms is described in detail in \cite{NZSHK23}. We used the open-source code of \citet{K22}, adapting it to the pressure-controlled internal flow considered in this work.

The 2D computational geometry was constructed in FEniCS 2019.1.0 \cite{FEniCS1,FEniCS2}. A conforming (or matching) mesh was generated using \texttt{mshr}, FEniCS's mesh generation tool. After meshing, the static reference domain was discretized into finite elements. We used on the order of $150,000$ to $200,000$ triangular elements in the unstructured mesh, having verified that this number ensures mesh independence of the results of a particular simulation. A mixed-function space was defined on the mesh using equal-order linear Lagrange elements, which is allowed since we will use stabilization.  A separate function space was used for the mesh displacement field in the mesh-motion problem. Both the mesh and the function spaces were defined on the static reference domain $\Omega_y$. We solved the nonlinear variational problems (both the fluid--structure problem and the mesh-motion problem) in FEniCS by defining a \texttt{NonlinearVariationalProblem()} and using the associated solver \texttt{NonlinearVariationalSolver()}, which employs Newton's method. The full technical details are described in \cite{K22}. 

Backward Euler (implicit) time-stepping was used with a small enough time step to ensure time-step independence of the solution. Within each time step, we (\textit{i}) solved the nonlinear variational problem associated with Eq.~\eqref{eq:fluid-solid residual} for velocity and pressure, holding the mesh deformation fixed, and (\textit{ii}) solved the nonlinear variational problem associated with the artificial mesh-motion problem for mesh displacement. We iterated these two steps within each time step until the fluid--structure residual (assembled to a vector) met a convergence criterion with a relative tolerance of $10^{-6}$ in the $L^2$ norm. This type of solution strategy is called ``quasi-direct'' coupling, meaning that the fluid--structure problem is solved as one block, while the mesh-motion problem is solved as a separate block, with inner iterations between the two.

For a typical simulation shown below, we time-stepped the transient solution forward for seven oscillation cycles, which we verified was sufficient to treat the solution as time-periodic. Specifically, cycle convergence was determined by ensuring that the maximum streaming (cycle-averaged) pressure difference over $Z$ between the sixth and seventh cycles was below $10^{-6}$.


\section{Results and discussion}
\label{sec:results}

In this section, we compare the theoretical predictions from Sec.~\ref{sec:2D} of the primary  (Sec.~\ref{sec:results_primary}) and secondary (Sec.~\ref{sec:results_secondary}) pressure and deformation profiles to the results of the finite element simulations described in Sec.~\ref{sec:simulations}.

Based on the principle of dynamic similarity, the simulations were set up and performed with geometric and material properties of the fluid and solid domains to match the seven corresponding dimensionless numbers in Table~\ref{table:param_2D_p_control}. The specific values of the numerous dimensional parameters in the ALE-FSI code are not relevant to list (they are, in a sense, partially arbitrary as long as the dimensionless numbers match), but they are available in the published data set \cite{RPC25_PURR}. In the simulations, $\Wor$ was varied by changing $\omega$ (with $h_0$, $\rho_f$, and $\mu_f$ fixed). For each $\Wor$, different values of $\gamma$ were considered by changing $\ell$ (with $\omega$, $C_W$, $\mu_f$, and $h_0$ fixed). Changing $\ell$ necessarily changes $\epsilon_{f,s}$ (from which $\theta$ and $\vartheta$ are computed for each simulation), which is why four values are listed for each in Table~\ref{table:param_2D_p_control}. In the discussion below, we state only the $\Wor$ and $\gamma$ values for each simulation, as they are the key parameters governing the physics discussed herein.

\subsection{Primary flow}
\label{sec:results_primary}

\subsubsection{Fluid pressure and velocity}

\begin{figure}[ht!]
     \centering
     \begin{subfigure}[b]{0.3\textwidth}
         \centering
         \includegraphics[width=\textwidth]{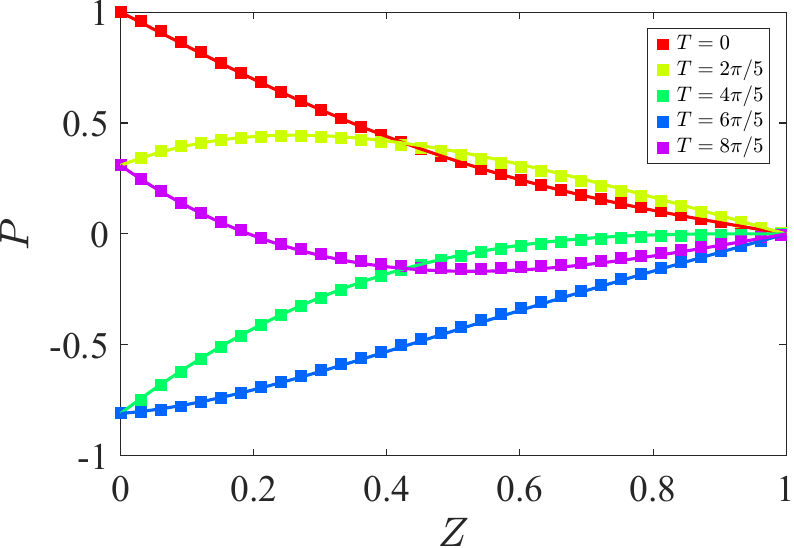}
         \caption{$\gamma=0.5,\ \Wor=1$}
     \end{subfigure}
     \hfill
     \begin{subfigure}[b]{0.3\textwidth}
         \centering
         \includegraphics[width=\textwidth]{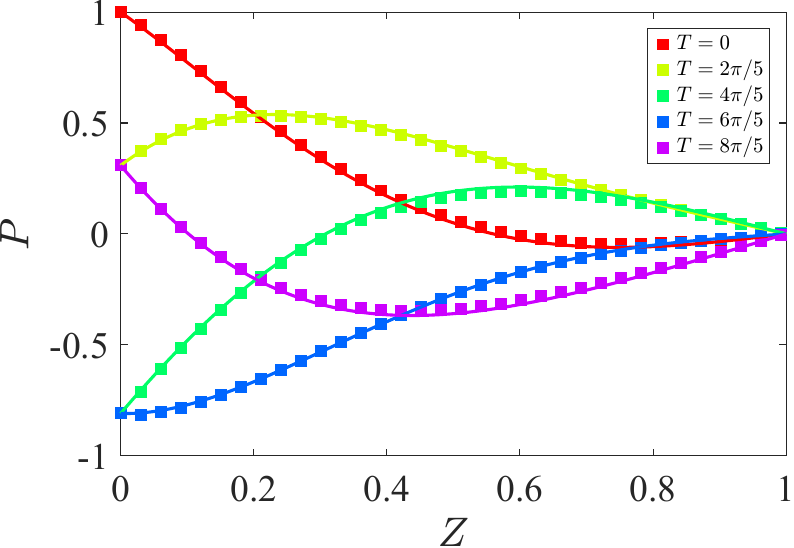}
         \caption{$\gamma=1.0,\ \Wor=2$}
     \end{subfigure}
     \hfill
     \begin{subfigure}[b]{0.3\textwidth}
         \centering
         \includegraphics[width=\textwidth]{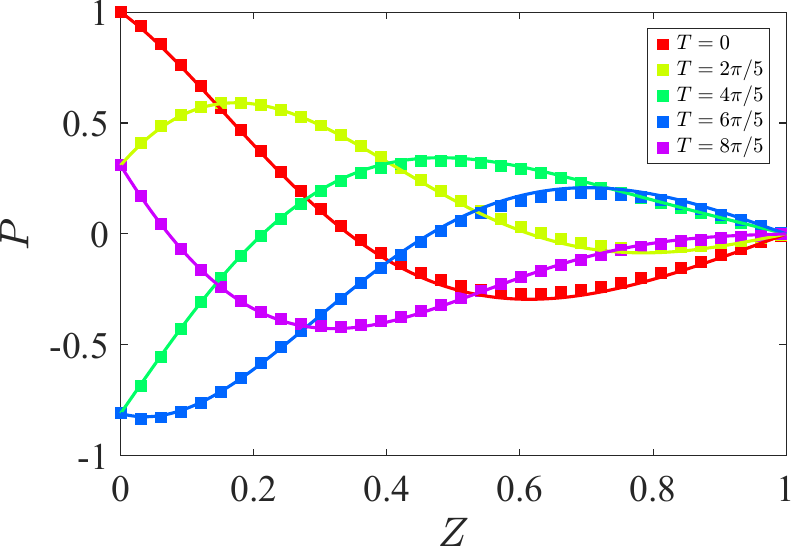}
         \caption{$\gamma=1.5,\ \Wor=3$}
     \end{subfigure}
     \caption{Pressure distribution along the fluidic channel over one flow oscillation cycle. As a periodic state has been established, we label the curves as $T=0,~2\pi/5,~4\pi/5,~6\pi/5,~8\pi/5$; $T=2\pi$ would overlap with $T=0$, so it is not shown. Solid curves represent the theoretical prediction $P_0(Z,T) = \Real[P_{0,a}(Z)\re^{\ri T}]$ obtained from Eq.~\eqref{eq:Pa0_soln}. Symbols represent the finite element simulation, from which the ``full'' pressure $P(Z,T)$ was obtained and plotted.}
     \label{fig:pressure curves}
\end{figure}

Figure~\ref{fig:pressure curves} presents the dimensionless pressure distribution, $P(Z,T)$, along the fluidic channel over one flow oscillation cycle for three representative value pairs of the elastoviscous number $\gamma$ and the Womersley number $\Wor$. The plots show a strong agreement (maximum root-mean-squared error $\approx0.006$ across all simulations shown) between the theoretical predictions, i.e., the primary pressure  $P_0(Z,T) = \Real[P_{0,a}(Z)\re^{\ri T}]$ obtained from Eq.~\eqref{eq:Pa0_soln} and the simulations, with the symbols overlapping the curves in most cases. This comparison demonstrates that the theory accurately captures the pressure oscillations despite $P_0$ being calculated from the linearization of the problem in $\beta$. That is to say, the coupling between the wall unsteadiness and flow unsteadiness, as represented by this leading-order solution, determines the nonuniform pressure gradient in the fluidic channel. From Fig.~\ref{fig:pressure curves}, we observe that, as $\Wor$ increases, the pressure becomes more oscillatory with $Z$, as suggested by the theory [recall Fig.~\ref{fig:kappa_effects}(a) and its discussion]. 

\begin{figure}[ht!]
     \centering
     \begin{subfigure}[b]{0.3\textwidth}
         \centering
         \includegraphics[width=\textwidth]{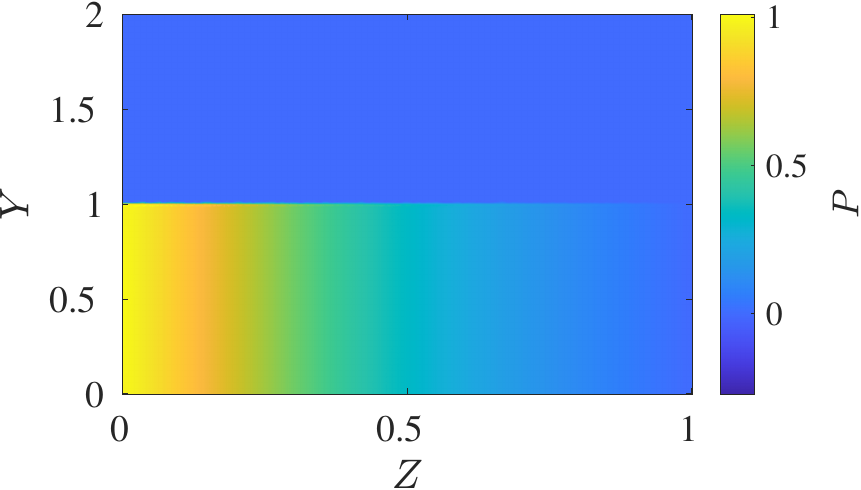}
         \caption{$\gamma=0.5,\ \Wor=1$}
     \end{subfigure}
     \hfill
     \begin{subfigure}[b]{0.3\textwidth}
         \centering
         \includegraphics[width=\textwidth]{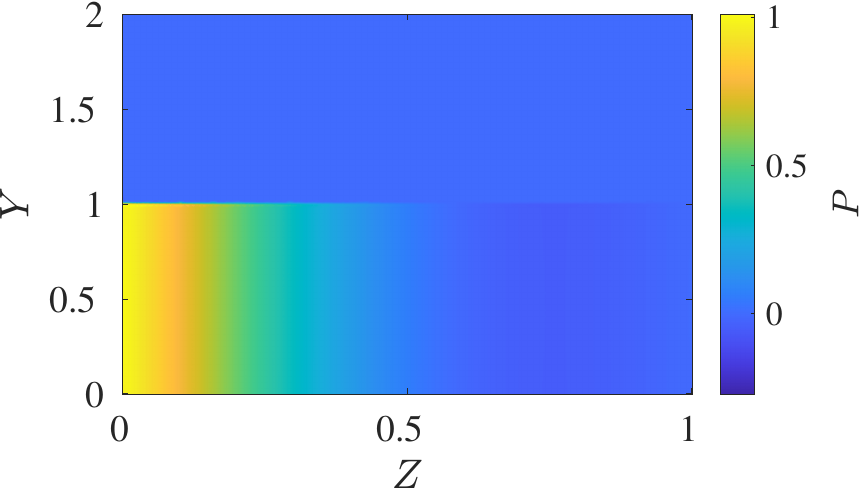}
         \caption{$\gamma=1.0,\ \Wor=2$}
     \end{subfigure}
     \hfill
     \begin{subfigure}[b]{0.3\textwidth}
         \centering
         \includegraphics[width=\textwidth]{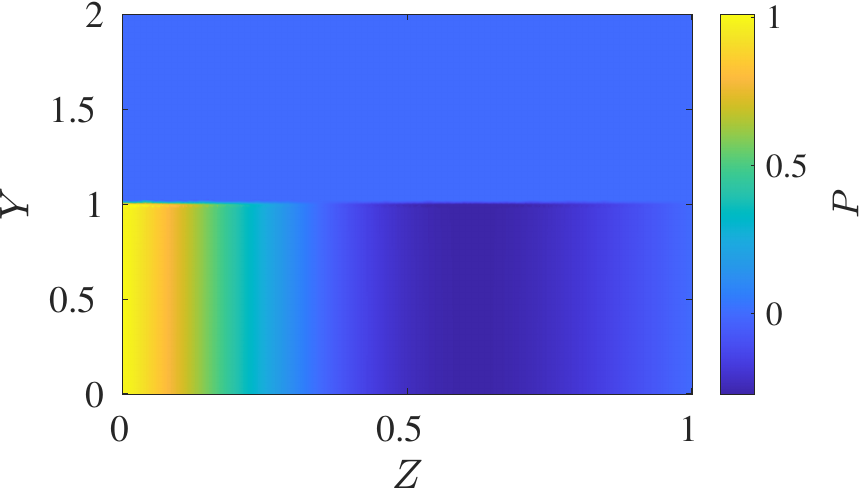}
         \caption{$\gamma=1.5,\ \Wor=3$}
     \end{subfigure}
     \caption{Visualization of the pressure field, $P$, obtained from simulations at the start of the cycle ($T=0$), depicting its variation in the $(Z,Y)$ plane. We set the pressure in the solid subdomain ($Y>1$) to zero, resulting in a uniform zero-pressure field throughout this subdomain.}
     \label{fig:contours pressure}
\end{figure}

\begin{figure}[ht!]
     \centering
     \begin{subfigure}[b]{0.3\textwidth}
         \centering
         \includegraphics[width=\textwidth]{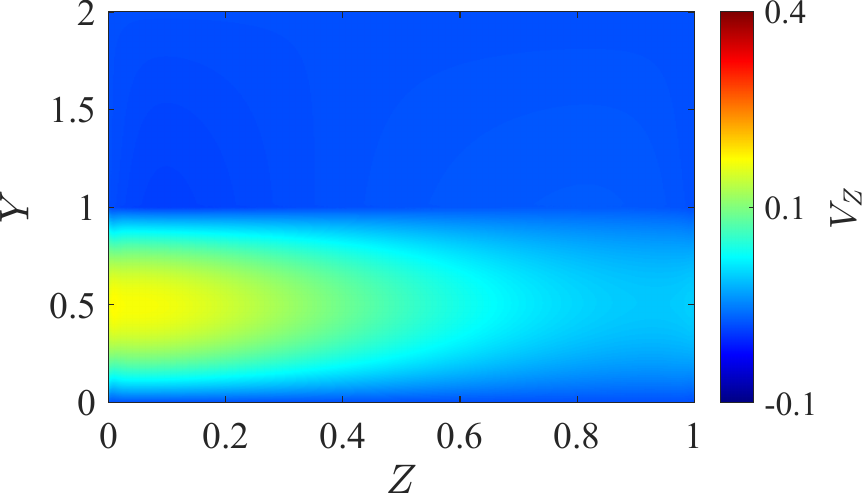}
         \caption{$\gamma=0.5,\ \Wor=1$}
     \end{subfigure}
     \hfill
     \begin{subfigure}[b]{0.3\textwidth}
         \centering
         \includegraphics[width=\textwidth]{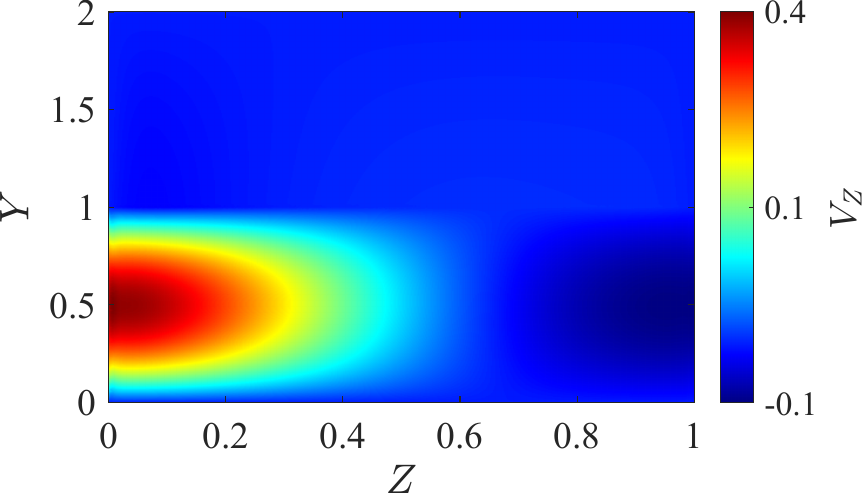}
         \caption{$\gamma=1.0,\ \Wor=2$}
     \end{subfigure}
     \hfill
     \begin{subfigure}[b]{0.3\textwidth}
         \centering
         \includegraphics[width=\textwidth]{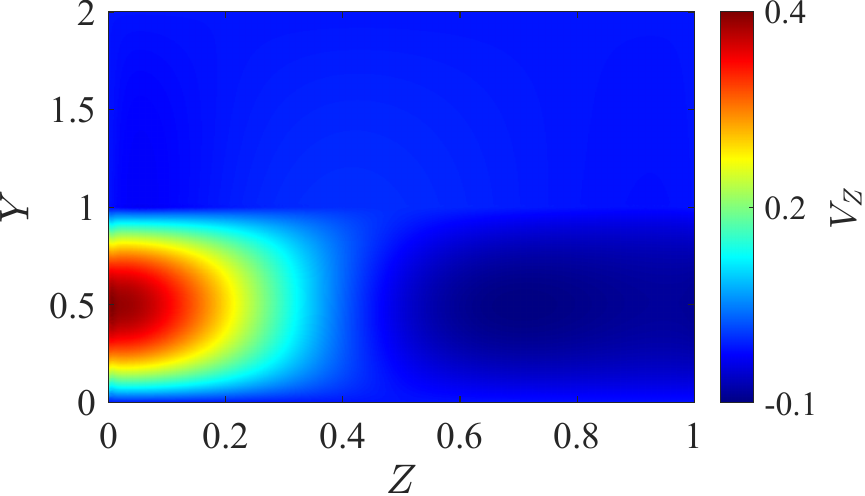}
         \caption{$\gamma=1.5,\ \Wor=3$}
     \end{subfigure}
     \caption{Visualization of the horizontal component of the velocity field, $V_Z$, obtained from simulations at the beginning of the cycle ($T=0$).}
     \label{fig:contours velocity}
\end{figure}

The pressure contour plots in Fig.~\ref{fig:contours pressure} verify the tenet of lubrication theory, namely that the pressure does not vary with $Y$ to the leading order in $\epsilon_f$. From these contour plots of $P$ in the $(Z,Y)$ plane at the start of a cycle, it is evident that the pressure is uniform vertically in the fluid subdomain, consistent with Eq.~\eqref{eq:ymom}. Additionally, the pressure attains negative values for $Z \gtrsim 0.34$ in Fig.~\ref{fig:contours pressure}(c), consistent with Fig.~\ref{fig:pressure curves}(c). In other words, the fluid exerts a pulling force on the wall surface in this part of the domain. 

Next, in Fig.~\ref{fig:contours velocity}, we show contour plots of the axial velocity $V_Z$ at the start of the cycle ($T=0$). As we progress from Fig.~\ref{fig:contours velocity}(a) to Fig.~\ref{fig:contours velocity}(c), the velocity's magnitude increases, as should be expected, since the elastoviscous number is larger. While the velocity profile looks parabolic for the lower $\Wor$ values, this is not the case for higher values. Indeed, the profile is already strongly blunted near the centerline in Fig.~\ref{fig:contours velocity}(c). This blunting is a manifestation of the oscillatory Stokes boundary layer, whose dimensionless thickness scales as $\Wor^{-1}$. In other words, for $\Wor \lesssim 1$ the viscous layer fills the entire channel height, and the profile remains quasi-parabolic, whereas for $\Wor \gtrsim 1$ the viscous layer is confined to a thin region near the walls, leaving a nearly plug-like core that is largely decoupled from the viscous stresses. 

Additionally, the velocity becomes increasingly localized near the inlet with higher $\Wor$ values, consistent with the localization of the pressure oscillations in Fig.~\ref{fig:pressure curves}(c). The increasing inlet-localization of the velocity field for $\Wor > 1$ is consistent with the spatial localization of the primary pressure oscillations [recall Figs.~\ref{fig:kappa_effects}(b) and \ref{fig:pressure curves}]. It is also worth noting that a velocity field exists in the solid subdomain, though its magnitude is significantly lower compared to that in the fluid subdomain, so it is barely perceptible in this figure.

\subsubsection{Elastic layer displacements}

\begin{figure}[t]
     \centering
     \begin{subfigure}[b]{0.3\textwidth}
         \centering
         \includegraphics[width=\textwidth]{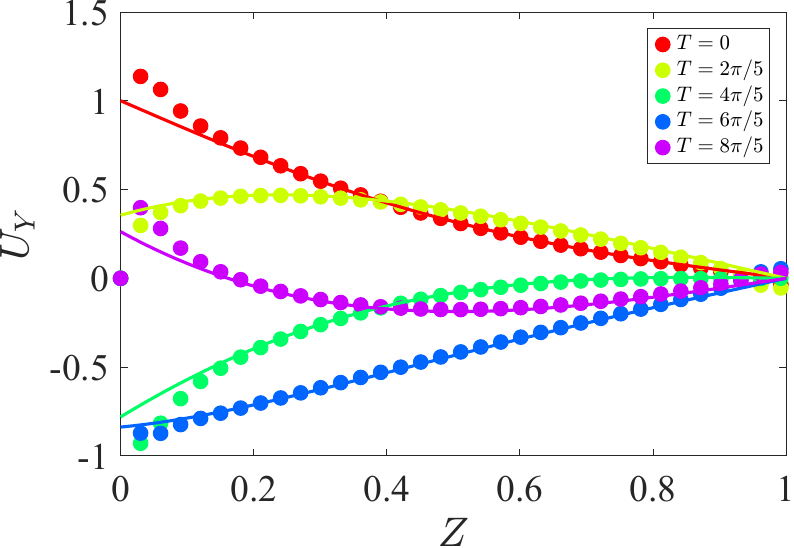}
         \caption{$\gamma=0.5,\ \Wor=1$}
     \end{subfigure}
     \hfill
     \begin{subfigure}[b]{0.3\textwidth}
         \centering
         \includegraphics[width=\textwidth]{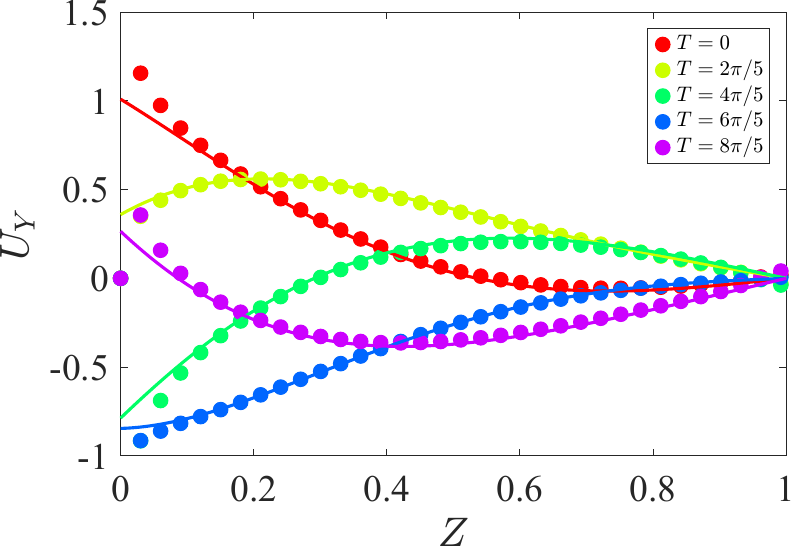}
         \caption{$\gamma=1.0,\ \Wor=2$}
     \end{subfigure}
     \hfill
     \begin{subfigure}[b]{0.3\textwidth}
         \centering
         \includegraphics[width=\textwidth]{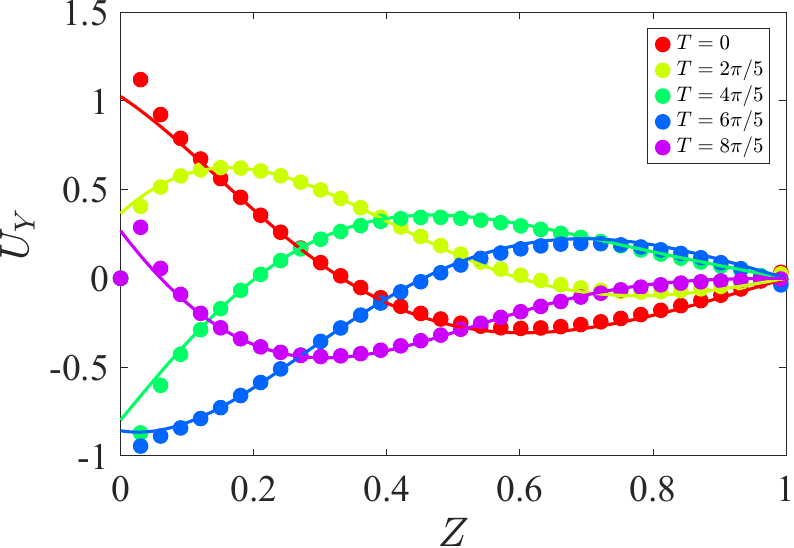}
         \caption{$\gamma=1.5,\ \Wor=3$}
     \end{subfigure}
     \caption{Vertical displacement of the fluid--solid interface. Solid curves represent the theoretical prediction $U_{Y,0}(Z,T) = \Real[U_{Y,0,a}(Z)\re^{\ri T}]$ obtained from Eqs.~\eqref{eq:Ua0_soln} and \eqref{eq:Pa0_soln}, while the symbols represent the vertical displacement component $U_Y$ obtained from the finite element simulation.}
     \label{fig:displacement for phases curves}
\end{figure}

In Fig.~\ref{fig:displacement for phases curves}, we present the vertical displacement of the fluid--solid interface, $U_Y(Z,T)$, over one flow oscillation cycle for the same three pairs of $\gamma$ and $\Wor$ as in the previous subsection. Since $U_Y$ at the interface is, to the leading order, directly determined by the pressure, the displacement curves are also ``wavier'' for larger values of $\gamma$ and $\Wor$. We observe that the combined foundation model, i.e., $U_{Y,0}(Z,T) = \Real[U_{Y,0,a}(Z)\re^{\ri T}]$ obtained from Eq.~\eqref{eq:UYa0_soln}, and the finite element simulations match, except near the inlet ($Z=0$). At the inlet, the combined foundation model~\eqref{eq:combined_foundation} does not respect the clamped (no displacement) boundary condition. However, we observe that this defect is highly localized and does not affect the displacement profile beyond the first $10-15\%$ of the fluidic domain's length. Clamping can be easily incorporated into the theory by regularizing the combined foundation model via weak tension, as shown in Appendix~\ref{app:clamping}. However, as shown in the latter, clamping cannot capture the pressure overshoot above $1$ seen in Fig.~\ref{fig:displacement for phases curves}, which is likely due to the presence of significant axial displacements, a point we address next.

\begin{figure}[t]
     \centering
     \begin{subfigure}[b]{0.3\textwidth}
         \centering
         \includegraphics[width=\textwidth]{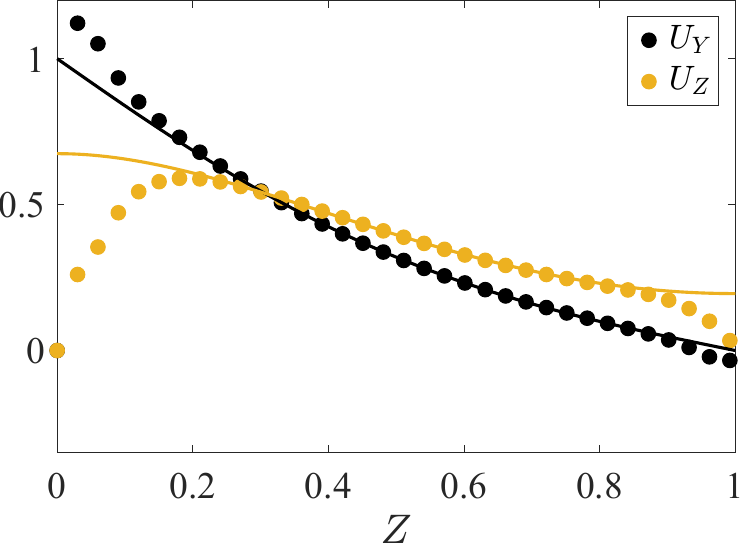}
         \caption{$\gamma=0.5,\ \Wor=1$}
     \end{subfigure}
     \hfill
     \begin{subfigure}[b]{0.3\textwidth}
         \centering
         \includegraphics[width=\textwidth]{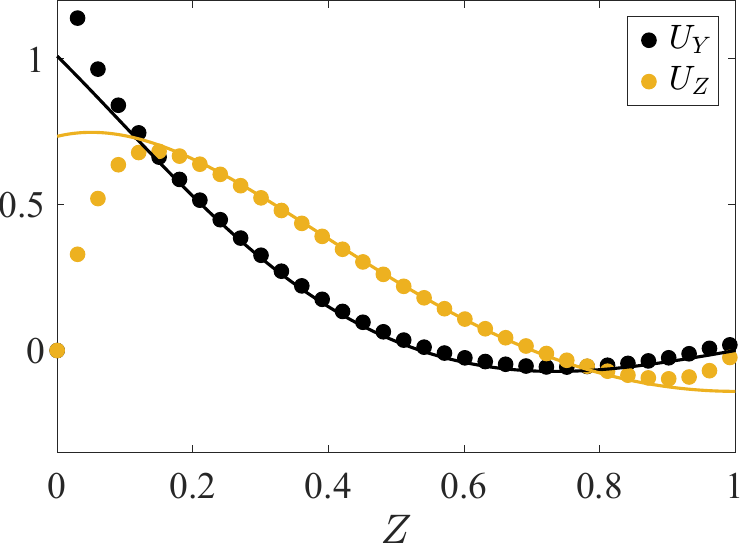}
         \caption{$\gamma=1.0,\ \Wor=2$}
     \end{subfigure}
     \hfill
     \begin{subfigure}[b]{0.3\textwidth}
         \centering
         \includegraphics[width=\textwidth]{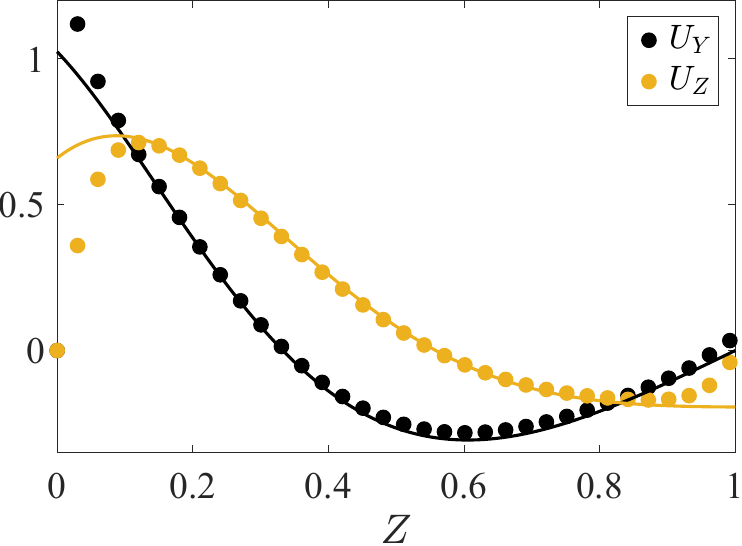}
         \caption{$\gamma=1.5,\ \Wor=3$}
     \end{subfigure}
     \caption{Displacement of the fluid--solid interface: the two components, $U_Y$ and $U_Z$, at the start of the cycle, i.e., $T=0$. Solid curves represent the theoretical predictions,  $U_{Y,0}(Z,T) = \Real[U_{Y,0,a}(Z)\re^{\ri T}]$ and $U_{Z,0}(Z,T) = \Real[U_{Z,0,a}(Z)\re^{\ri T}]$ obtained from Eqs.~\eqref{eq:Ua0_soln} and \eqref{eq:Pa0_soln}, while the symbols represent the finite element simulation. Both displacement components have been scaled the same way, per Eq.~\eqref{eq:ndvars}.}
     \label{fig:displacement curves}
\end{figure}

Since the elastic layer is nearly incompressible, the vertical displacement $U_Y$ engenders a horizontal displacement $U_Z$ at the interface, by the Poisson effect, which is readily seen at the start of the cycle in Fig.~\ref{fig:displacement curves} and in agreement (maximum root-mean-squared error $\approx0.007$ across all simulations shown) with the theoretical predictions from Eqs.~\eqref{eq:UZa0_soln} and \eqref{eq:Pa0_soln}. This holds true throughout the entire cycle. Notice that if we had resorted to the Winkler foundation model, i.e., $\theta=\vartheta=0$ in Eqs.~\eqref{eq:combined_foundation}, then we would have been unable to explain the observed nonzero horizontal displacement. However, in the simulations, the elastic wall is clamped on the left side of the channel (i.e., $U_Y=U_Z=0$ exactly at $Z=0$), which leads to a localized disagreement with the theory near the inlet (since no inlet conditions can be specified for the displacement under the foundation model). This effect is most pronounced at the beginning/end of the cycle when the displacements at the inlet are largest, as can be seen in Fig.~\ref{fig:displacement for phases curves}. Finally, note that both displacement components have been scaled the same way, per Eq.~\eqref{eq:ndvars} and \cite{CV20}. However, one should keep in mind that the $z$ scale is much longer than the $y$ scale, thus the characteristic strain $\sim u_z/\ell \ll u_y/h_0$.

\begin{figure}[ht]
     \centering
     \begin{subfigure}[b]{0.3\textwidth}
         \centering
         \includegraphics[width=\textwidth]{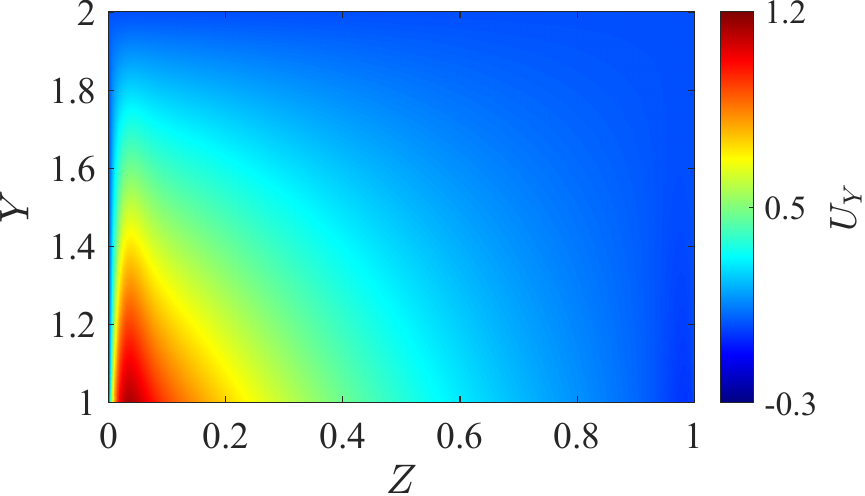}
         \caption{$\gamma=0.5,\ \Wor=1$}
     \end{subfigure}
     \hfill
     \begin{subfigure}[b]{0.3\textwidth}
         \centering
         \includegraphics[width=\textwidth]{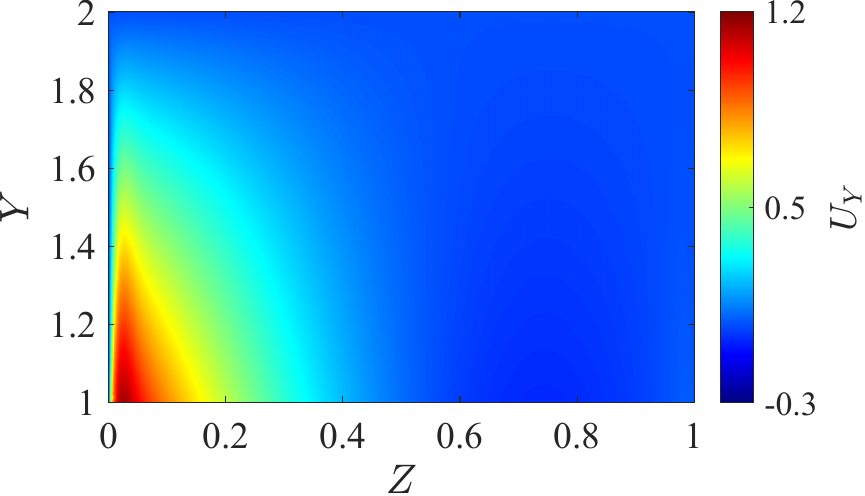}
         \caption{$\gamma=1.0,\ \Wor=2$}
     \end{subfigure}
     \hfill
     \begin{subfigure}[b]{0.3\textwidth}
         \centering
         \includegraphics[width=\textwidth]{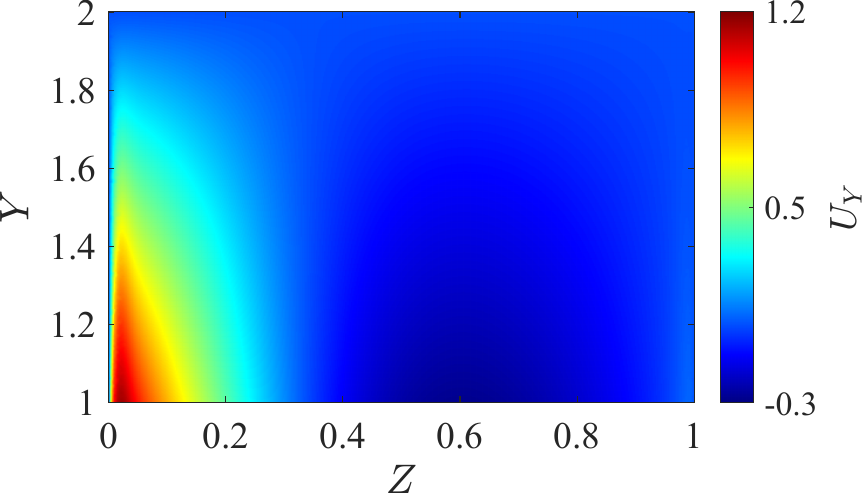}
         \caption{$\gamma=1.5,\ \Wor=3$}
         \label{fig:contour vert disp for gamma=1.5 wo=3}
     \end{subfigure}\\[15pt]
    \begin{subfigure}[b]{0.3\textwidth}
         \centering
         \includegraphics[width=\textwidth]{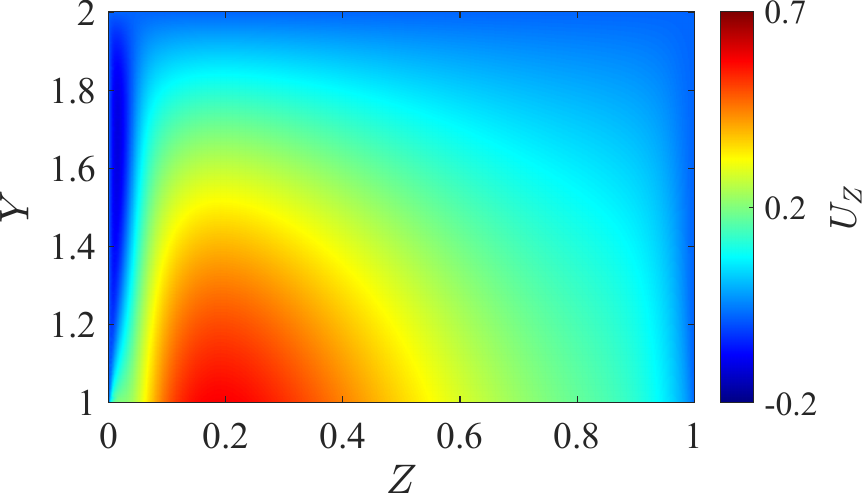}
         \caption{$\gamma=0.5,\ \Wor=1$}
     \end{subfigure}
     \hfill
     \begin{subfigure}[b]{0.3\textwidth}
         \centering
         \includegraphics[width=\textwidth]{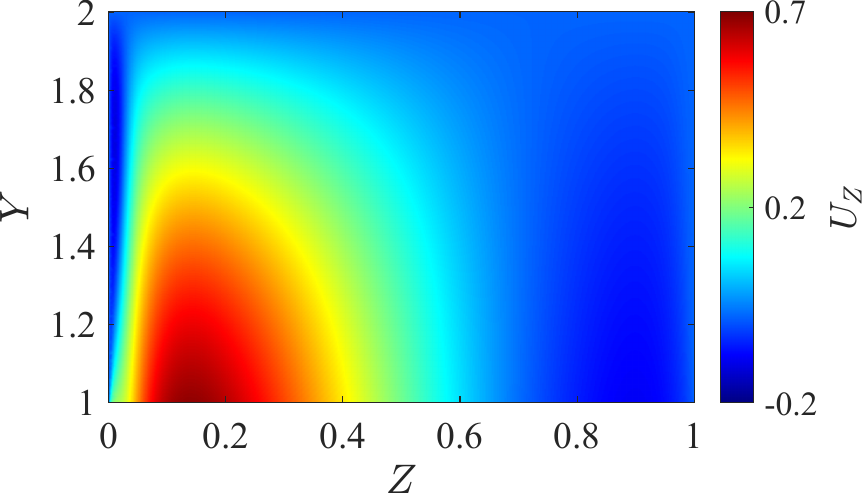}
         \caption{$\gamma=1.0,\ \Wor=2$}
     \end{subfigure}
     \hfill
     \begin{subfigure}[b]{0.3\textwidth}
         \centering
         \includegraphics[width=\textwidth]{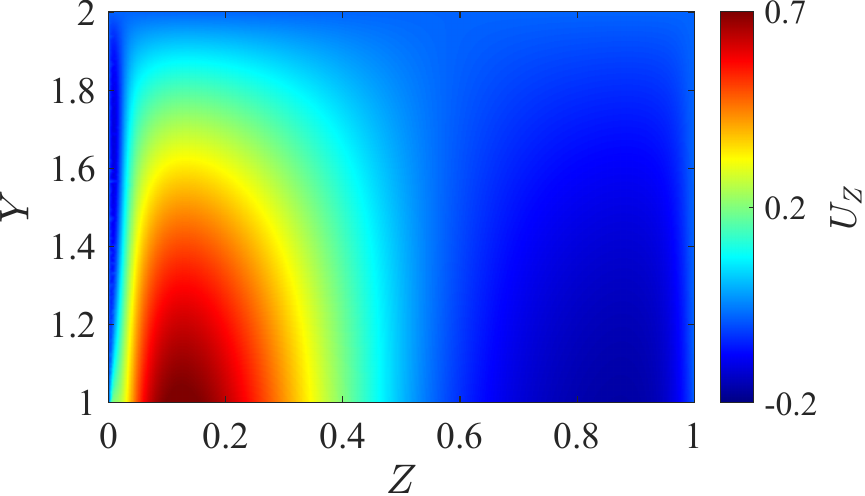}
         \caption{$\gamma=1.5,\ \Wor=3$}
     \end{subfigure}
     \caption{Visualization of (a,b,c) the vertical component of the displacement field, $U_Y$, and (d,e,f) the horizontal component of the displacement field, $U_Z$, obtained from simulations at the beginning of the cycle ($T=0$) in the solid subdomain.}
     \label{fig:contours disp}
\end{figure}

Figures~\ref{fig:displacement for phases curves} and \ref{fig:displacement curves} provide information solely on the displacement at the fluid--solid interface. To gain a deeper understanding of the displacement field within the solid subdomain, we visualize the simulated displacement field, both $U_Y$ and $U_Z$ components, in Fig.~\ref{fig:contours disp} at the beginning of the cycle, in the $(Z,Y)$ plane using contour plots. Both displacement field components are zero at the constrained rigid boundaries (along $Z=0$, $Z=1$, and $Y=2$), recall Figs.~\ref{fig:schematic} and \ref{fig:2DDomain}. The largest displacements develop near the fluid--solid interface ($Y=1$) and the inlet ($Z=0$), where the pressure is largest. 

We observe that the vertical displacement [Fig.~\ref{fig:contours disp}(a,b,c)] is much more localized than the horizontal displacement [Fig.~\ref{fig:contours disp}(d,e,f)]. This observation is a direct consequence of the near-incompressibility of the confined elastic layer. The vertical displacement $U_Y$ is driven locally by the pressure load, which decays rapidly with Z (particularly at higher $\Wor$, recall Fig.~\ref{fig:pressure curves}), so it remains concentrated near the high-pressure inlet. The horizontal displacement $U_Z$, however, must spread over a longer piece of the layer to accommodate the volume-preserving constraint (Poisson effect) while satisfying the zero-displacement conditions at both the inlet and outlet, producing the broadly distributed pattern visible in Fig.~\ref{fig:contours disp}(d,e,f). The combined foundation model~\eqref{eq:combined_foundation} captures these key features of the fluid--solid \emph{interface}'s displacement (recall Figs.~\ref{fig:displacement for phases curves} and \ref{fig:displacement curves}), which is the key piece of physics needed to understand the primary pressure oscillations, and next the secondary (cycle-averaged) streaming characteristics.

\subsection{Secondary (cycle-averaged) flow}
\label{sec:results_secondary}

\subsubsection{Fluid pressure}

A nonzero cycle-averaged (streaming) pressure $\langle P_1 \rangle$ arises because both the flow-induced wall (geometric nonlinearity) deformation and the advective inertia (flow nonlinearity) lead to cycle-averages of products of two oscillatory quantities that, while individually having zero mean, yield a nonzero mean over the oscillation cycle, as shown by Eq.~\eqref{eq:P1_avg}. Both contributions are important, as emphasized by \citet{ZR24}, in determining the streaming pressure profiles.

\begin{figure}
     \centering
     \begin{subfigure}[b]{0.3\textwidth}
         \centering
         \includegraphics[width=\textwidth]{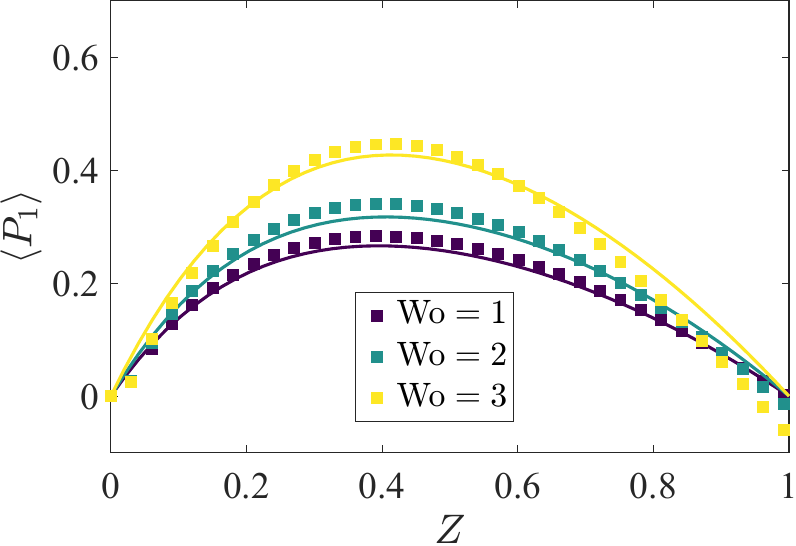}
         \caption{$\gamma=0.5$}
     \end{subfigure}
     \hfill
     \begin{subfigure}[b]{0.3\textwidth}
         \centering
         \includegraphics[width=\textwidth]{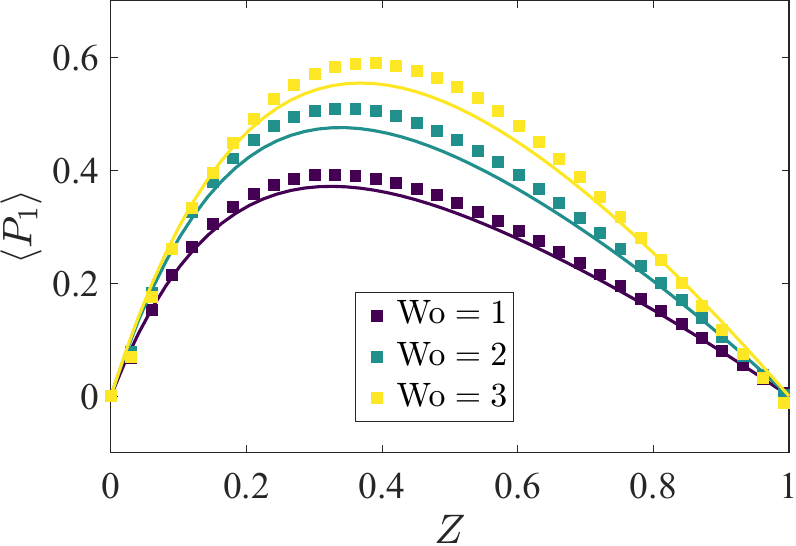}
         \caption{$\gamma=1.0$}
     \end{subfigure}
     \hfill
     \begin{subfigure}[b]{0.3\textwidth}
         \centering
         \includegraphics[width=\textwidth]{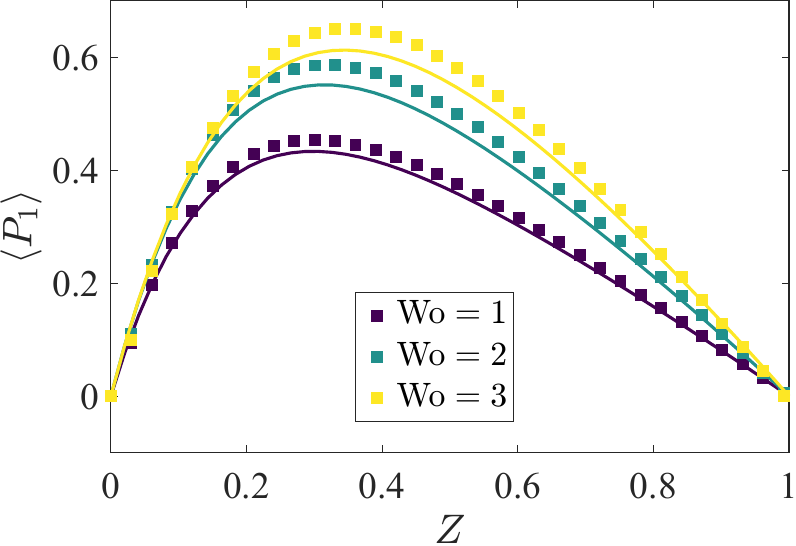}
         \caption{$\gamma=1.5$}
     \end{subfigure}
     \caption{Streaming (cycle-averaged) pressure distribution, $\langle P_1 \rangle \equiv \langle P \rangle/\beta$, along the fluidic channel for three different values of the elastoviscous number $\gamma$, and three different values of the Womersley number $\Wor$ each. Solid curves represent the theory, namely Eqs.~\eqref{eq:Q1_avg} and \eqref{eq:P1_avg}, while the symbols represent the values obtained from the finite element simulations.}
     \label{fig:streaming pressure curves}
\end{figure}

To demonstrate this effect, Fig.~\ref{fig:streaming pressure curves} shows the comparison of theory and simulation. It can be seen that the magnitude of the streaming pressure grows with $\gamma$. Additionally, the peak streaming pressure shifts towards the inlet of the channel, and the curves exhibit increased asymmetry, with the peak shifting away from $Z=0.5$. Overall, the elastoinertial rectification theory developed in Sec.~\ref{sec:Obeta_prob} is successful in capturing, qualitatively and quantitatively (maximum root-mean-squared error $\approx0.03$ across all simulations shown), the physics at hand. The simulations likewise successfully resolve this higher-order (weak) effect. The resonance-like behavior of the streaming pressure and flow rate as $\Wor$ is varied was discussed in Sec.~\ref {sec:Obeta_prob} and already demonstrated by the equally good agreement between the theory and simulations in Fig.~\ref{fig:theoretical time-averaged-flux}(a).

In Fig.~\ref{fig:streaming pressure curves}(a), we do observe some artifacts near the outlet at the largest $\Wor$ value, which can be resolved by choosing a longer channel [as in Fig.~\ref{fig:streaming pressure curves}(b,c)]. However, for each subpanel in Fig.~\ref{fig:streaming pressure curves}, we chose to maintain the same length for each simulation. Figure~\ref{fig:streaming pressure curves} also suggests that there is a slight but growing discrepancy between the simulation and the theory as $\Wor$ increases. 

\subsubsection{Fluid--solid interface displacements}

\begin{figure}
     \centering
         \begin{subfigure}[b]{0.3\textwidth}
         \centering
         \includegraphics[width=\textwidth]{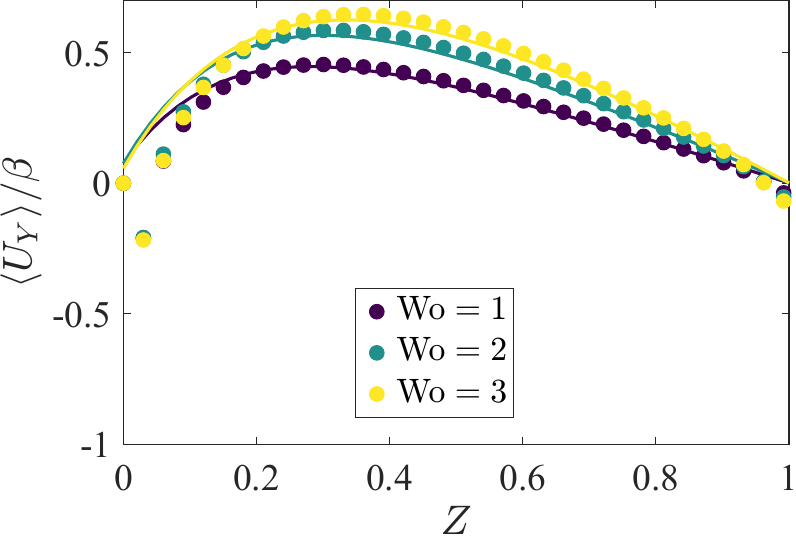}
         \caption{}
     \end{subfigure}
     \qquad\quad
     \begin{subfigure}[b]{0.3\textwidth}
         \centering
         \includegraphics[width=\textwidth]{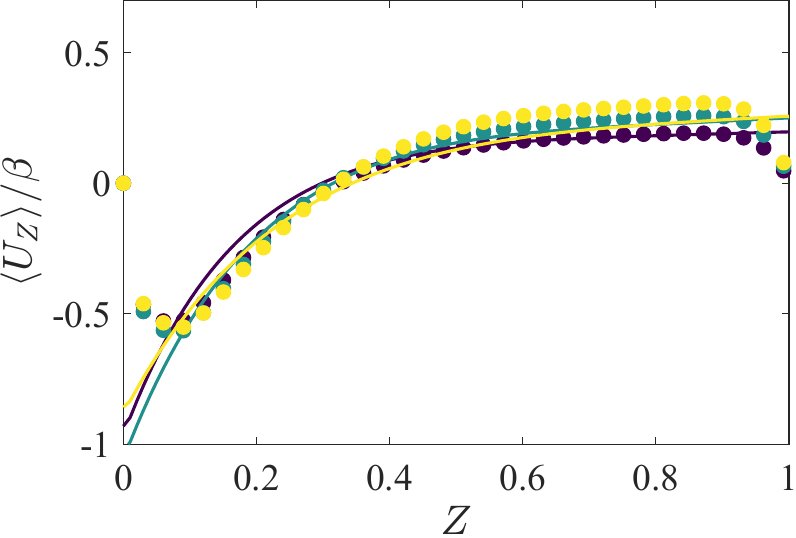}
         \caption{}
     \end{subfigure}
     \caption{Streaming (cycle-averaged) (a) vertical and (b) horizontal displacements along the fluid--solid interface for $\gamma = 1.5$ and three different values of $\Wor$. Solid curves represent the theoretical solutions, Eqs.~\eqref{eq:combined_foundation_avg}, evaluated using  Eqs.~\eqref{eq:Q1_avg} and \eqref{eq:P1_avg}, while the symbols represent the values obtained from the finite element simulations.}
     \label{fig:time-avg disp curves}
\end{figure}

Figure~\ref{fig:time-avg disp curves} presents the scaled cycle-averaged displacements, 
\begin{subequations}\label{eq:combined_foundation_avg}
    \begin{align}
        {\langle U_Y \rangle}/{\beta} &= \langle P_1 \rangle - (\theta - \vartheta\mathfrak{f}_1) \frac{d^2\langle P_1 \rangle}{ dZ^2},\\
        {\langle U_Z\rangle}/{\beta} &=\left(\frac{\epsilon_s{h_0}}{C_W G}\mathfrak{f}_1 - \epsilon_s \hat{\vartheta}\right) \frac{d\langle P_1\rangle}{dZ},
    \end{align}%
\end{subequations}
obtained from Eqs.~\eqref{eq:combined_foundation}, for the representative value of $\gamma = 1.5$ and the three values of $\Wor$. The requisite derivatives of $\langle P_1 \rangle$ are computed by second-order finite difference from the numerically computed streaming pressure above.

Similarly to the case of the primary displacements, $U_{Y,0}$ and $U_{Z,0}$ in Fig.~\ref{fig:displacement curves}, the theory and simulations generally agree well also for $\langle U_Y \rangle$ and $\langle U_Z \rangle$ in Fig.~\ref{fig:time-avg disp curves}, save for edge effects, which are more prominent at this order of approximation. In particular, in Fig.~\ref{fig:time-avg disp curves}(a), the central shape of the $\langle U_Y \rangle/\beta$ profiles, their magnitudes, and dependence on $\Wor$ are fully captured by the elastoinertial rectification theory, i.e., Eqs.~\eqref{eq:combined_foundation_avg}.

Both averaged displacement components shown in Fig.~\ref{fig:time-avg disp curves} increase with $\Wor$, similar to the streaming pressure in Fig.~\ref{fig:streaming pressure curves}. While the trend is captured well for $\langle U_Y \rangle / \beta$ in Fig.~\ref{fig:time-avg disp curves}(a), this is only the case for $Z\gtrsim 0.5$ for $\langle U_Z \rangle / \beta$ in Fig.~\ref{fig:time-avg disp curves}(b), though the discrepancies are small (maximum root-mean-squared error $\approx0.02$ across all simulations shown). Importantly, from Fig.~\ref{fig:time-avg disp curves}, we learn that $\langle U_Z \rangle/\beta$ is comparable to $\langle U_Y \rangle/\beta$, which further highlights the need for the use of the combined foundation model (over a simpler Winkler foundation that presupposes $U_Z \equiv 0$). Consequently, despite the viscous shear forces in the flow being significantly weaker than the pressure forces (due to the small aspect ratio of the geometry), it does not follow that the viscous shear on the fluid--solid interface produces negligible axial displacement of a confined, nearly-incompressible 2D elastic layer.

Finally, we emphasize that these nontrivial $\mathrm{O}(\beta)$ displacements are a new feature compared to previous works on the subject. Like the streaming pressure, the streaming displacements are determined entirely by the primary flow in the theory. The fact that the $\mathrm{O}(\beta)$ displacements agree well with the simulations (Fig.~\ref{fig:time-avg disp curves}) provides new insights into the combined foundation model and validates its assumptions; such a comparison is not readily available in the literature. Finally, we emphasize that these are 2D effects, as, for example, in a 3D elastoinertial problem \cite{HPFC25}, there is no significant $\langle U_Z \rangle$ because the channel wall can deform in the perpendicular (transverse) $X$ direction.

\section{Conclusion}
\label{sec:conclusion}

To address a knowledge gap in the literature on oscillatory flows in compliant confinements, we presented a theoretical and computational investigation of a two-way-coupled oscillatory flow of a Newtonian fluid in a 2D initially rectangular channel with a confined elastic layer serving as its top wall. Specifically, we employed the combined foundation model developed by \citet{CV20} as a reduced model of the deformation of the 2D elastic layer. We coupled the combined foundation model to the fluid flow via the kinematic boundary condition at the fluid--solid interface. Additionally, we scaled the fluid's axial momentum equation in the lubrication limit in such a way as to retain advective inertia, which contributes to the secondary flow for weak compliance, as shown by \citet{ZR24}. From the reduced governing equations, assuming weak channel compliance but without restricting the oscillation frequency, we derived expressions for the primary and secondary cycle-averaged (streaming) pressure profiles and the flow-rate enhancement. To validate our assumptions and theoretical results, we performed ALE-FSI simulations using a solver implemented in FEniCS and compared the pressure and displacement profiles predicted by the theory with those from the simulations. 

Our key finding is good agreement between theory and simulations, as indicated by low root-mean-squared errors, for both the primary and secondary (streaming) pressure profiles across a range of the key dimensionless numbers, namely the Womersley $\Wor$ and elastoviscous $\gamma$ numbers. This agreement validates the combined foundation model as the appropriate description of the interaction between an oscillatory flow and a nearly incompressible 2D elastic layer, and confirms that both the geometric nonlinearity of the deforming fluid domain and the fluid's advective nonlinearity must be retained to accurately predict the streaming pressure. Interestingly, unlike previous work \cite{IWC20} and conventional wisdom, which suggest that shear stresses are much weaker than normal stresses at the fluid--solid interface, the combined foundation model \cite{CV20} predicts significant axial displacements driven by both the wall shear stress and pressure gradient when the elastic layer thickness is similar to the fluidic channel height. Finding good agreement for the primary displacements, too, we demonstrated this feature of the problem. Additionally, these benchmarks indicated that inlet clamping effects, which are neglected in the theory but retained in the simulations, have a negligible overall effect on the pressure characteristics. 

Nevertheless, we found that systematic edge effects arise from the boundary conditions on the confined elastic layer, which are captured by the ALE-FSI simulations but not the model. As \citet{CV20} noted, ``if the imposed boundary conditions are incompatible with the displacement field suggested... we expect a boundary layer,'' on the order of the thickness of the elastic layer, ``to resolve this incompatibility.'' Indeed, our clamped boundary conditions at the inlet and outlet are incompatible, though they lead to only localized discrepancies. Accounting for the boundary-layer structure near the clamped edges and properly matching to the corner singularities \citep{CL77_2} presents a challenging but rewarding avenue for future work. However, an additional important finding in our study is that disagreements arising from the elastic problem's boundary conditions do not significantly affect the streaming pressure profiles, which are set by the primary pressure and vertical displacement. Therefore, our work demonstrates the phenomenon of \emph{elastoinertial} rectification in a 2D channel and provides a predictive theory for it, extending the original work on axisymmetric geometries \cite{ZR24}.

Furthermore, the theory shows how the shape of the primary pressure profile is determined by both $\Wor$ and $\gamma$. Interestingly, the combined foundation model \cite{CV20}, unique to nearly-incompressible 2D layers considered herein and distinct from the deformation model for 3D axisymmetric geometries \cite{ZR24} and 3D channels \cite{HPFC25}, shows the existence of values of the Womersley number for which the primary pressure exhibits large axial oscillations, followed by a cutoff. Similarly, but for different values of the Womersley number, our theory predicts augmentation of the secondary (streaming) pressure's magnitude; recall the discussion of Fig.~\ref{fig:theoretical time-averaged-flux}(a) at the end of Sec.~\ref{sec:2D}. These resonance-like behaviors are absent in 3D channels \cite{HPFC25}, or in 3D axisymmetric geometries \cite{ZR24}, and are a unique feature of the combined foundation model, specific to nearly incompressible 2D layers. Importantly, in microfluidic systems, vertical confinement of the layer and the lack of significant out-of-plane strains lead to substantial axial displacements that require a mathematical description, beyond the Winkler-like mechanism previously used for 3D Cartesian and 3D axisymmetric geometries. Therefore, the 2D problem analyzed herein serves as a canonical model that further elucidates the physics of elastoinertial rectification beyond previous work. Our results may be useful for optimizing and designing compliant microfluidic systems, in which electro-osmotic mechanisms have previously been shown to yield similar resonance-like effects \cite{C06,SDDC22}. 

In future work, it would be of interest to extend the present analysis to non-Newtonian fluids---specifically, inelastic shear-thinning \cite{PN22} 
or viscoelastic \cite{Asghari2020} fluids relevant to oscillatory microfluidic flows \cite{DDS20,Muduetal24}. It may also be of interest to consider secondary flows in closed conduits with elastic boundaries that confine \emph{compressible} fluids \cite{O25}. Another interesting problem is elastoinertial rectification in  periodically loaded 2D poroelastic materials, such as those studied by \citet{FPM23,FPM25}. It may also be relevant to better understand the effect of the viscoelasticity of the soft confining layer \cite{AC20,KB25} on the streaming flow, which could be done using the deformation models of \citet{PKVS16}, or by deriving a new viscoelastic combined foundation model in the spirit of \citet{CV20}.


\begin{acknowledgments}
This research was partially supported by the U.S.\ National Science Foundation under Grant No.\ CMMI-2245343. We thank B.\ Rallabandi for many insightful discussions on elastoinertial rectification. We also thank D.\ Kamensky for answering questions regarding the technical and computational implementation of the ALE-FSI solver in FEniCS.
\end{acknowledgments}

\smallskip

U.M.R.\ performed the simulations and led the software development, data curation, visualization, and validation, contributed equally to the investigation, and supported the formal analysis and the writing of the original draft.
S.D.P.\ contributed equally to the investigation and formal analysis, performed further simulations for the revision, and supported the validation, visualization, and the writing of the original draft and its revision.
I.C.C.\ conceptualized the work, administered and supervised the project, acquired funding, led the development of the methodology, contributed equally to the formal analysis, supported the investigation, visualization, and validation, as well as software development for the revision, and led the writing of the original draft and its revision.

\section*{Data availability}
The data and codes that support the computational findings of this article are openly available on the Purdue University Research Repository at \url{https://dx.doi.org/10.4231/BWA1-E343} \cite{RPC25_PURR}.


\bibliography{references,other_refs}



\appendix

\section{Analytical approximation for the advective term and asymptotic expressions}
\label{app:analytical_approximations}

Obtaining the streaming pressure and flow rate from numerical solutions of Eq.~\eqref{eq:Vz1T_avg_pde} may appear inelegant to some. To remedy the situation, \citet{ZR24} introduced a procedure for the numerical approximation of the contributions arising from the advective terms. Adapting their approach to the channel geometry, we replace the axial velocity~\eqref{eq:Vza0_soln} with a parabolic profile that matches their centerline velocity. Specifically,
\begin{equation}
    {V}_{Z,0}(Y,Z) \approx 
    \underbrace{\frac{1}{\ri{\Wor}^2}\left[1-\frac{1}{\cos\left(\ri^{3/2} {\Wor}/2 \right)}\right]}_{\mathfrak{v}_0(\Wor)} 4Y(1-Y) \left( -\frac{d P_{0}}{d Z} \right).
    \label{eq:Vz0_approx}
\end{equation}
Using Eqs.~\eqref{eq:Vz0_approx} and \eqref{eq:com}, we find the approximate vertical component of the velocity field:
\begin{equation}
     {V}_{Y,0}(Y,Z) \approx \mathfrak{v}_0(\Wor)\left(\frac{4Y^3}{3}-2Y^2\right) \left(-\frac{d^2P_0}{d{Z}^2}\right).
     \label{eq:Vy0_approx}
\end{equation}
Substituting Eqs.~\eqref{eq:Vz0_approx} and \eqref{eq:Vy0_approx} into Eq.~\eqref{eq:Vz1T_avg_pde}, we find an approximate expression for $\widetilde{\langle V_{Z,1}\rangle}$:
\begin{equation}
    \widetilde{\langle V_{Z,1}\rangle}(Y,Z) \approx \frac{\Wor^2}{\gamma} \left\langle \left(\mathfrak{v}_0 \frac{dP_0}{dZ} \right) \left(\mathfrak{v}_0\frac{d^2P_0}{dZ^2}\right) \right\rangle \frac{2}{45} \left(-7Y + 15Y^4 - 12Y^5 + 4Y^6\right),
\end{equation}
and
\begin{equation}
    \int_0^1 \widetilde{\langle V_{Z,1} \rangle} \, dY   
    \approx -\frac{3\Wor^2}{35\gamma} \bigg\langle \left(\mathfrak{v}_0 \frac{dP_0}{dZ} \right) \left(\mathfrak{v}_0\kappa^2 P_0\right) \bigg\rangle,
    \label{eq:int_til_Vz1_approx}
\end{equation}
having used Eq.~\eqref{eq:Pa0_ode} to replace ${d^2P_0}/{dZ^2}$ by $\kappa^2 P_0$.
Meanwhile, no approximation is made in computing the contribution of the effective slip velocity. Using Eqs.~\eqref{eq:Ua0_soln} and \eqref{eq:dVz0dYat1}, we find
\begin{equation}
    -\frac{1}{2}\left.\left\langle U_{Y,0} \frac{\partial V_{Z,0}}{\partial Y} \right|_{Y=1}\right\rangle = - \frac{1}{2} \left\langle \left(\frac{\mathfrak{f}_0 }{\ri \gamma} \kappa^2 P_0 \right) \left[ \frac{\ri^{1/2}}{\Wor}\tan\left(\ri^{3/2} \Wor/2\right) \left(- \frac{d P_0}{d Z} \right) \right] \right\rangle .
    \label{eq:avg_U0dVz0dY_exact}
\end{equation}
Then, 
\begin{equation}
    -\frac{1}{2}\left.\left\langle U_{Y,0} \frac{\partial V_{Z,0}}{\partial Y} \right|_{Y=1}\right\rangle + \int_0^1 \widetilde{\langle V_{Z,1} \rangle} \, dY \\
    \approx 
    -\frac{1}{2} \Real\Bigg\{ \underbrace{\left[ \frac{\ri^{1/2}}{\Wor}\tan\left(\ri^{3/2} \Wor/2\right) \frac{\mathfrak{f}_0^*}{2\ri} + \frac{3\Wor^2}{35}|\mathfrak{v}_0|^2 \right] \frac{(\kappa^2)^*}{\gamma}}_{\mathfrak{C}(\Wor,\gamma)} P_{0,a}^* \frac{d P_{0,a}}{d Z} \Bigg\}.
    \label{eq:combined_approx}
\end{equation}
Now, approximate versions of  Eqs.~\eqref{eq:Q1_avg} and \eqref{eq:P1_avg} [or Eqs.~\eqref{eq:Q1_avg_q_control} and \eqref{eq:P1_avg_q_control}] could be derived using Eq.~\eqref{eq:combined_approx}.

For the pressure-controlled regime, using Eq.~\eqref{eq:Pa0_soln}, we evaluate
\begin{multline}
    \int_0^Z P_{0,a}^* \frac{dP_{0,a}}{dZ}  \, dZ 
    = -\frac{1}{4} \frac{\kappa}{ |\sinh\kappa|^2} \left[ \frac{\cosh\big(2\Real[\kappa]\big) - \cosh\big(2(1-Z)\Real[\kappa]\big)}{\Real[\kappa]} \right. \\
    \left. + \ri \frac{\cos\big(2\Imag[\kappa]\big) - \cos\big(2(1-Z)\Imag[\kappa]\big)}{\Imag[\kappa]} \right] .
    \label{eq:int_P0as_dP0aDz}
\end{multline}
    
\subsection{Quasi-rigid limit (\texorpdfstring{$\gamma\to0$}{gamma->0})}
\label{app:approx_gm0}

As in \cite{PWC23,ZR24}, it is of interest to consider the leading contributions of the flow oscillations in the limit of negligible viscous--elastic interaction, which in the present context is the limit $\gamma\to0$, keeping leading terms in $\Wor \ll 1$. From Eq.~\eqref{eq:int_P0as_dP0aDz},
\begin{equation}
    \int_0^Z P_{0,a}^* \frac{dP_{0,a}}{dZ}  \, dZ 
    \sim \frac{1}{2} Z(Z-2), \qquad \gamma\to0.
    \label{eq:approx4}
\end{equation}
Then, using \textsc{Mathematica} to expand $\mathfrak{C}(\Wor,\gamma)$ from Eq.~\eqref{eq:combined_approx}, and substituting that result and Eq.~\eqref{eq:approx4} into Eq.~\eqref{eq:Q1_avg}, we find
\begin{equation}
    \langle Q_1 \rangle \approx \frac{1}{16} \left( 1 - \frac{31}{2100} \Wor^4 \right) + \mathrm{O}(\Wor^6),
    \qquad \gamma\rightarrow0,
\end{equation}
where the leading $1/16$ term agrees with \cite{PWC23} but the coefficient of $\Wor^4$ has been updated by inertia here. Similarly, from Eq.~\eqref{eq:P1_avg},
\begin{equation}
    \langle P_1 \rangle \approx 12 Z (1-Z) \langle Q_1\rangle,
    \qquad \gamma\rightarrow0,
\end{equation}
where the leading $\tfrac{3}{4} Z(1-Z)$ term agrees with \cite{PWC23} but the coefficient of $\Wor^4$ has been updated by inertia here.

\subsection{Low-frequency limit (\texorpdfstring{$\Wor\to0$}{Wo->0})}
\label{app:approx_W0}

It is of interest to consider the leading contribution of the viscous--elastic interaction in the limit of negligible flow oscillations, which in the present context is the limit $\Wor\to0$, keeping leading terms in $\gamma\ll1$. From Eq.~\eqref{eq:int_P0as_dP0aDz},
\begin{equation}
    \int_0^Z P_{0,a}^* \frac{dP_{0,a}}{dZ}  \, dZ 
    \sim  \frac{1}{2} Z(Z-2)+\ri\gamma Z(Z-2)[2+Z(Z-2)], \qquad \Wor\to0.
    \label{eq:approx5}
\end{equation}
Thus, using \textsc{Mathematica} to expand $\mathfrak{C}(\Wor,\gamma)$ from Eq.~\eqref{eq:combined_approx}, and substituting that result and Eq.~\eqref{eq:approx5} into Eq.~\eqref{eq:Q1_avg}, we find
\begin{equation}
    \langle Q_1 \rangle \approx \frac{1}{16} \left\{ 1-\left[12(2\theta+\vartheta) + 36(2\theta+\vartheta)^2\right]\gamma^2 \right\} + \mathrm{O}(\gamma^4),
    \qquad \Wor\rightarrow0.
\end{equation}


\section{Flow-rate-controlled regime}
\label{app:flow_rate_control}

In the flow-rate-controlled regime, the definitions of the Womersley and elastoviscous numbers remain the same, as they are independent of the pressure scale. Letting $q_0$ be the imposed amplitude of the flow rate oscillations (per unit width), the axial velocity scale is now $v_c=q_0/h_0$, whence the pressure scale is $p_c=\mu q_0/(\epsilon_f h_0^2)$. Thus, the compliance number, based on this $p_c$ rather than $p_0$, is $\beta=\mathcal{C}_W\mu q_0/(\epsilon_f h_0^3)$.

Next, from Eq.~\eqref{eq:Qa0_soln}, we recognize that imposing the leading-order flow rate requires setting a boundary condition on the pressure gradient at the inlet. Thus, for the flow-rate-controlled regime, the primary problem's governing equations from Sec.~\ref{sec:O1_prob} remain the same, the only change is in the first boundary condition in Eq.~\eqref{eq:P0_bc_a}, which becomes ${dP_{0,a}}/{dZ} = -{1}/{\mathfrak{f}_0(\Wor)}$ at $Z=0$. The solution is easily found to be
\begin{equation}
    P_{0,a}(Z) = \frac{\sinh\big(\kappa(1-Z)\big)}{\mathfrak{f}_0(\Wor) \kappa \cosh \kappa},
    \label{eq:Pa0_soln_q_control}
\end{equation}
where $\kappa$ is given in Eq.~\eqref{eq:Pa0_soln}, and $U_{Y,0,a}(Z)$ is still given by Eq.~\eqref{eq:Ua0_soln}, now in terms of $P_{0,a}(Z)$ from Eq.~\eqref{eq:Pa0_soln_q_control}.

The cycle-averaged governing equations at $\mathrm{O}(\beta)$ are still Eqs.~\eqref{eq:O2_avg_prob} from Sec.~\ref{sec:Obeta_prob}, but the boundary conditions now reflect that the ``full'' flow rate was imposed on the leading-order problem:
\begin{subequations}\begin{empheq}[left = \empheqlbrace]{alignat=2}
    \langle V_{Z,1} \rangle|_{Y=0} &= 0,\qquad & \langle V_{Z,1} \rangle |_{Y=1} &= - \langle U_{Y,0} (\partial V_{Z,0}/\partial Y) |_{Y=1} \rangle, \label{eq:Vz1_avg_bc_q_control}\\
   \langle Q_1 \rangle|_{Z=0} &=0, & \langle P_1 \rangle|_{Z=1} &= 0. \label{eq:P1_avg_bc_q_control}
\end{empheq}\label{eq:O2_avg_bc_q_control}\end{subequations}
The solution for $\langle V_{Z,1}\rangle$ is again found according to Eqs.~\eqref{eq:Vz1T} and \eqref{eq:Vz1T_avg_pde}. 
However, now the inlet BC in Eq.~\eqref{eq:P1_avg_bc_q_control} and the continuity equation~\eqref{eq:Q1_avg_pde} at $\mathrm{O}(\beta)$ dictate that 
\begin{equation}
    \langle Q_1 \rangle = \int_0^1 \langle V_{Z,1} \rangle \, dY   
    = -\frac{1}{12}\frac{d \langle P_1\rangle }{d Z} - \frac{1}{2}\left.\left\langle U_{Y,0} \frac{\partial V_{Z,0}}{\partial Y} \right|_{Y=1}\right\rangle + \int_0^1 \widetilde{\langle V_{Z,1} \rangle} \, dY \underbrace{=}_{\text{by Eq.~\eqref{eq:P1_avg_bc_q_control}}} 0,
    \label{eq:Q1_avg_q_control}
\end{equation}
from which we can immediately solve for the streaming pressure
\begin{equation}
    \langle P_1 \rangle(Z) = -12 \int_{Z}^{1} \left[  -\frac{1}{2}\left.\left\langle U_{Y,0} \frac{\partial V_{Z,0}}{\partial Y} \right|_{Y=1}\right\rangle + \int_0^1 \widetilde{\langle V_{Z,1} \rangle} \, dY \right] dZ ,
    \label{eq:P1_avg_q_control}
\end{equation}
having applied the outlet BC in Eq.~\eqref{eq:P1_avg_bc_q_control}.

Finally, we note that when comparing to experiments (in which there is less control over the actual inlet and outlet conditions than there is in simulations), there are other possible BC combinations as well \cite{HPFC25}.


\section{Primary pressure and displacement solutions incorporating clamping at both ends}
\label{app:clamping}

As the fluid--solid interface deforms, it must elongate, especially near the high-pressure inlet of the fluidic channel, which causes tension. The tension can be considered constant (independent of $Z$) because, within the lubrication approximation, the shear stresses on the interface are weaker than the pressure forces \cite{HBDB14,KGV17}. Thus, the combined foundation model~\eqref{eq:combined_foundation_Y} can be ``regularized'' by introducing weak tension \cite{LP96,WC22} proportional to the linearized interface curvature, $d^2 U_{Y,0}/dZ^2$. This modification leads to an exactly solvable primary problem, as in Sec.~\ref{sec:O1_prob}.

The coupled ODEs governing the primary pressure and displacement solutions under the combined foundation model regularized by weak tension, specifically Eqs.~\eqref{eq:U0_pde_a} and \eqref{eq:Q0_pde_a}, having eliminated $Q_{0,a}$ via Eq.~\eqref{eq:Qa0_soln}, are now
\begin{subequations}\begin{empheq}[left = \empheqlbrace]{align}
    \left[\theta-\vartheta \mathfrak{f}_1(\Wor)\right] \frac{d^2 P_{0,a}}{d Z^2}-\mathcal{T}\frac{d^2 U_{Y,0,a}}{d Z^2}+U_{Y,0,a}&=P_{0,a}, 
    \label{eq:Pa0_ode_CV_tension}\\
    -\mathfrak{f}_0(\Wor)\frac{d^2 P_{0,a}}{d Z^2}+\ri \gamma U_{Y,0,a}&=0, \label{eq:Ua0_ode_CV_tension}
\end{empheq}\label{eq:ODEs_CV_tension}%
\end{subequations}
where $\mathcal{T} > 0$ is a tension coefficient. A detailed discussion of how to model the dimensionless tension coefficient $\mathcal{T}$ can be found in \cite{WC22,PWC23}. 
Equations~\eqref{eq:ODEs_CV_tension} are subject to the BCs:
\begin{subequations}\begin{empheq}[left = \empheqlbrace]{alignat=2}
    U_{Y,0,a}(0)&=0 ,\qquad &U_{Y,0,a}(1)&=0,\label{eq:BCs_U_tension}\\
    P_{0,a}(0)&=1 \text{ \textbf{OR} } \frac{dP_{0,a}}{dZ}(0)= -\frac{1}{\mathfrak{f}_0(\Wor)} ,\qquad &P_{0,a}(1)&=0.\label{eq:BCs_P_tension}
\end{empheq}\label{eq:BCs_tension}%
\end{subequations}
Eliminating $U_{Y,0,a}$ between the two equations in \eqref{eq:ODEs_CV_tension}, we obtain a fourth-order ODE for $P_{0,a}$:
\begin{equation}
    -\mathcal{T}\frac{\mathfrak{f}_0(\Wor)}{\ri\gamma}\frac{d^4 P_{0,a}}{d Z^4} + \left[\left[\theta-\vartheta \mathfrak{f}_1(\Wor)\right] + \frac{\mathfrak{f}_0(\Wor)}{\ri\gamma}\right]\frac{d^2 P_{0,a}}{d Z^2} - P_{0,a} = 0.
    \label{eq:Pa0_ode_CV_tension_2}
\end{equation}
Using Eq.~\eqref{eq:Ua0_ode_CV_tension}, the displacement BCs~\eqref{eq:BCs_U_tension} can be rewritten as two further pressure BCs:
\begin{equation}
    \frac{d^2 P_{0,a}}{d Z^2}(0)=\frac{d^2 P_{0,a}}{d Z^2}(1) = 0.
    \label{eq:BCs_P_tension_2}
\end{equation}

The characteristic equation of the linear ODE~\eqref{eq:Pa0_ode_CV_tension_2} is 
\begin{equation}
    \mathcal{T}\mathfrak{f}_0 \lambda^4 - \left( \ri\gamma\left[\theta-\vartheta \mathfrak{f}_1(\Wor)\right] + \mathfrak{f}_0\right) \lambda^2 + \ri\gamma = 0,
\end{equation}
and the four roots (solutions) $\{\pm\lambda_1,\pm\lambda_2\}$ are easily found (though the expressions are lengthy). Thus, the general solution of Eq.~\eqref{eq:Pa0_ode_CV_tension_2} is
\begin{equation}
    P_{0,a}(Z) = a_{1,1} \cosh(\lambda_1 Z) + a_{1,2} \sinh(\lambda_1 Z) + a_{2,1} \cosh(\lambda_2 Z) + a_{2,2} \sinh(\lambda_2 Z).
    \label{eq:Pa0_gen_sol_CV_tension}
\end{equation}
Applying the four BCs from Eqs.~\eqref{eq:BCs_P_tension} and \eqref{eq:BCs_P_tension_2} to the general solution in Eq.~\eqref{eq:Pa0_gen_sol_CV_tension}, we obtain a linear system for the coefficients:
\begin{subequations}\begin{empheq}[left = \empheqlbrace]{align}
    \lambda_1^2 a_{1,1} + \lambda_2^2 a_{2,1} &= 0,\\
    \lambda_1^2 \left( a_{1,1} \cosh\lambda_1 + a_{1,2} \sinh\lambda_1 \right) + \lambda_2^2 \left( a_{2,1} \cosh\lambda_2 + a_{2,2} \sinh\lambda_2\right) &=0,\\
    a_{1,1} + a_{2,1} = 1 \text{ \textbf{OR} } \lambda_1 a_{1,2} + \lambda_2 a_{2,2} &= -1/\mathfrak{f}_0(\Wor), \label{eq:constants_inlet_P_BC}\\
    a_{1,1} \cosh\lambda_1 + a_{1,2} \sinh\lambda_1 + a_{2,1} \cosh\lambda_2 + a_{2,2} \sinh\lambda_2 &= 0.
\end{empheq}\end{subequations}
Solving this system, we find
\begin{equation}
    a_{2,1} = - \left(\frac{\lambda_1}{\lambda_2}\right)^2 a_{1,1},\qquad
    a_{2,2} = \left(\frac{\lambda_1}{\lambda_2}\right)^2 a_{1,1} \coth \lambda_2,\qquad
    a_{1,2} = -a_{1,1} \coth \lambda_1.
\end{equation}

Next, after some trigonometric simplification:
\begin{align}
    P_{0,a}(Z) &= a_{1,1} \left[  \frac{\sinh\big(\lambda_1 (1-Z)\big)}{\sinh \lambda_1} - \left(\frac{\lambda_1}{\lambda_2}\right)^2 \frac{\sinh\big(\lambda_2 (1-Z)\big)}{\sinh \lambda_2} \right], \label{eq:P0a_soln_CV_clamped}\\
    U_{Y,0,a}(Z) &= a_{1,1} \lambda_1^2 \frac{\mathfrak{f}_0(\Wor)}{\ri\gamma} \left[  \frac{\sinh\big(\lambda_1 (1-Z)\big)}{\sinh \lambda_1} -  \frac{\sinh\big(\lambda_2 (1-Z)\big)}{\sinh \lambda_2} \right]. \label{eq:U0a_soln_CV_clamped}
\end{align}

Finally, $a_{1,1}$ is determined from the remaining  equation~\eqref{eq:constants_inlet_P_BC} (corresponding to the inlet pressure BC):
\begin{equation}
    a_{1,1} = 
    \begin{cases}
    \displaystyle    
    \frac{\lambda_2^2}{\lambda_2^2-\lambda_1^2},
    &\qquad\text{pressure control},\\[10pt]
    \displaystyle
    \frac{1}{\mathfrak{f}_0(\Wor)} \left(\frac{\lambda_2^2}{\lambda_2^2\lambda_1\coth\lambda_1 -\lambda_1^2\lambda_2\coth\lambda_2}\right), &\qquad\text{flow rate control}.
    \end{cases}
\label{eq:a11}%
\end{equation}

Note that the limit of $\mathcal{T}\to0$ is a singular one as the highest derivative in the ODE~\eqref{eq:Pa0_ode_CV_tension_2} is lost. However, the pressure phasor's amplitude ODE then obviously reduces to Eq.~\eqref{eq:Pa0_ode}, which we solved above.

The $\mathrm{O}(\beta)$ problem (Sec.~\ref{sec:Obeta_prob}) remains unchanged, and the streaming quantities can be determined using Eqs.~\eqref{eq:P0a_soln_CV_clamped} and \eqref{eq:U0a_soln_CV_clamped}.

\begin{figure}
     \centering
         \begin{subfigure}[b]{0.4\textwidth}
         \centering
         \includegraphics[width=\textwidth]{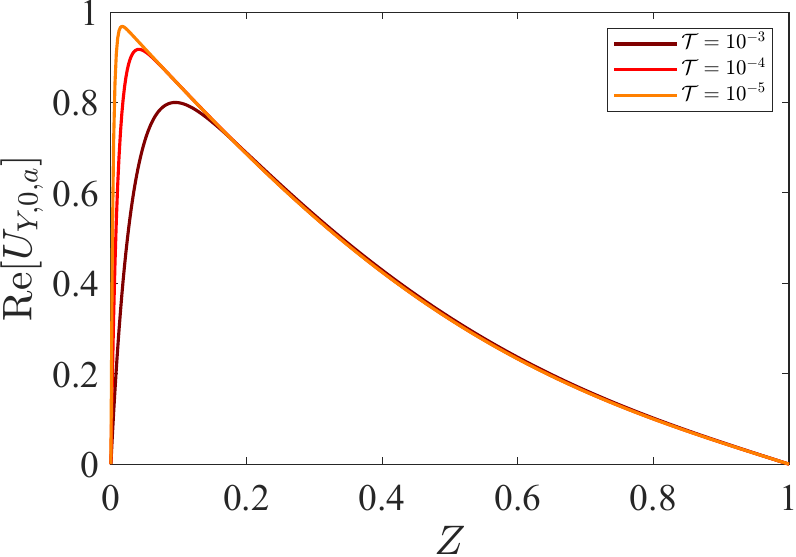}
         \caption{}
     \end{subfigure}
     \qquad\quad
     \begin{subfigure}[b]{0.4\textwidth}
         \centering
         \includegraphics[width=\textwidth]{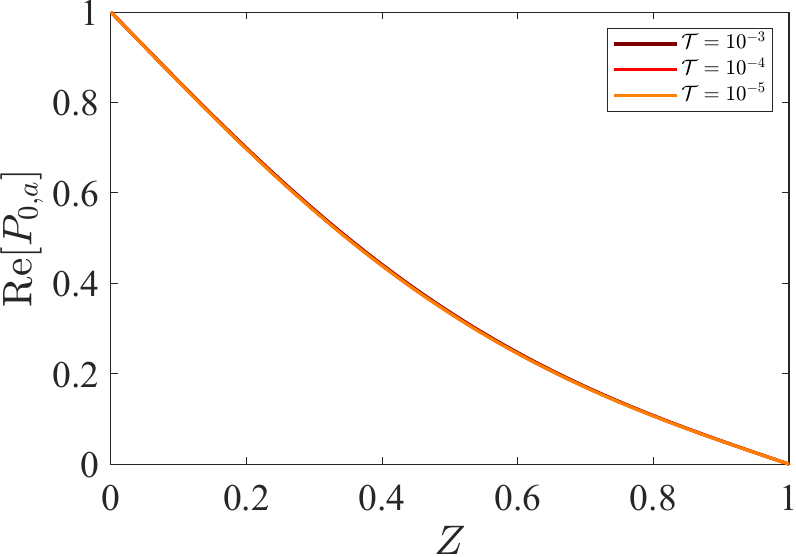}
         \caption{}
     \end{subfigure}
     \caption{The (a) vertical displacement of the fluid--solid interface $\mathrm{Re}[U_{Y,0,a}]$ and (b) the pressure $\mathrm{Re}[P_{0,a}]$, evaluated from Eqs.~\eqref{eq:U0a_soln_CV_clamped} and \eqref{eq:P0a_soln_CV_clamped}, respectively, for three different values of the tension coefficient $\mathcal{T}$, with $\gamma=0.5$ and $\Wor=1$. Note that the three curves in (b) overlap.}
     \label{fig:tension}
\end{figure}

To assess whether incorporating tension into the model could explain the overshoot near the inlet, observed in Figs.~\ref{fig:displacement for phases curves} and \ref{fig:displacement curves} above, in Fig.~\ref{fig:tension} we plot the vertical displacement component at the fluid--solid interface for three small values of $\mathcal{T}$, based on the solution from this Appendix, namely Eqs.~\eqref{eq:P0a_soln_CV_clamped} and \eqref{eq:U0a_soln_CV_clamped} for the values of $\gamma$ and $\Wor$ considered in Fig.~\ref{fig:displacement curves}(a). As illustrated by Fig.~\ref{fig:tension}(a), enforcing clamping at $Z=0$ leads to a peak in $\Real[U_{Y,0,a}]$. The peak shifts closer to $Z=0$ and a magnitude of $1$ as $\mathcal{T}\to0^+$. However, the peak never exceeds $1$. In fact, it is evident that weak tension ($\mathcal{T}\ll1$ but $\ne0$) causes an \emph{undershoot} in the vertical displacement. In Fig.~\ref{fig:tension}(b), we demonstrate that the leading-order pressure, $\mathrm{Re}[P_{0,a}]$, is mostly unaffected by weak axial tension caused by the clamping BC.


\end{document}